\documentclass[
    reprint,
    nofootinbib,
    amsmath,amssymb,
    aps,
    pra,
    floatfix,
    superscriptaddress
]{revtex4-2}

\usepackage[utf8]{inputenc}

\usepackage[colorlinks,citecolor=blue,linkcolor=blue,urlcolor=blue]{hyperref}

\usepackage{mathtools}
\usepackage{url}
\usepackage[title]{appendix}
\usepackage{braket}
\usepackage{slashed}
\usepackage{bbm}
\usepackage{xcolor}
\usepackage{pgfplots}
\usepackage{upgreek}
\usepackage{xspace}
\usepackage{graphicx}
\usepackage{epstopdf}

\usepackage[pdf]{pstricks}

\pgfplotsset{compat=1.18}

\usepackage{cleveref}
\Crefname{equation}{Eq.}{Eqs.}
\Crefname{figure}{Fig.}{Figs.}
\Crefname{tabular}{Tab.}{Tabs.}
\Crefname{section}{Sec.}{Secs.}

\Crefmultiformat{equation}{Eqs.~(#2#1#3)}{ and~(#2#1#3)}{,~(#2#1#3)}{ and~(#2#1#3)}
\let\cref\undefined

\newcommand{\I}{\mathrm{i}}
\newcommand{\E}{\mathrm{e}}
\newcommand{\dd}{\mathrm{d}}

\newcommand{\of}[1]{%
    \mathchoice
        {\!\left({#1}\right)}%
        {\!\left({#1}\right)}%
        {\left({#1}\right)}%
        {\left({#1}\right)}%
}
\newcommand{\ofb}[1]{%
    \mathchoice
        {\!\left[{#1}\right]}%
        {\!\left[{#1}\right]}%
        {\left[{#1}\right]}%
        {\left[{#1}\right]}%
}

\DeclareMathOperator{\cov}{cov}

\newcommand{\suoo}{$\mathrm{SU}\!\left(1,1\right)$\xspace}
\newcommand{\eqdef}{\overset{\mathrm{def.}}{=}}
\DeclareMathOperator{\auto}{auto}
\DeclareMathOperator{\cross}{cross}

\makeatletter
\renewcommand*\env@matrix[1][*\c@MaxMatrixCols c]{%
  \hskip -\arraycolsep
  \let\@ifnextchar\new@ifnextchar
  \array{#1}}
\makeatother

\usepackage[caption=false]{subfig}
\usepackage{siunitx}

\newcommand{\El}{\mathcal{L}}

\renewcommand{\AA}{\mathcal{A}}
\newcommand{\BB}{\mathcal{B}}
\newcommand{\CC}{\mathcal{C}}

\newcommand{\NN}{\mathcal{N}}

\newcommand{\KK}{\mathcal{K}}

\newcommand{\Eta}{\mathrm{H}}
\newcommand{\Beta}{\mathrm{B}}

\counterwithin*{equation}{section}
\renewcommand{\theequation}{\arabic{section}.\arabic{equation}}
\let\appx\appendix
\renewcommand{\appendix}{\appx\renewcommand{\theequation}{\Alph{section}.\arabic{equation}}}

\usepackage{placeins}

\usepackage{tabularx}
\newcolumntype{Y}{>{\centering\arraybackslash}X}

\graphicspath{{./figures/sketches/}}

\begin{document}

    \title{Phase sensitivity of spatially broadband high-gain SU(1,1) interferometers}
    
    \author{D.~Scharwald}
        \affiliation{Department of Physics, Paderborn University, Warburger Straße 100, D-33098 Paderborn,  Germany}
    \author{T.~Meier}
        \affiliation{Department of Physics, Paderborn University, Warburger Straße 100, D-33098 Paderborn,  Germany}
        \affiliation{Institute for Photonic Quantum Systems (PhoQS), Paderborn University, Warburger Straße 100, D-33098 Paderborn, Germany}
    \author{P.~R.~Sharapova}
        \affiliation{Department of Physics, Paderborn University, Warburger Straße 100, D-33098 Paderborn, Germany}

    \begin{abstract}
        Nonlinear interferometers are promising tools for quantum metrology, as they are
        characterized by an improved phase sensitivity scaling compared to
        linear interferometers operating with classical light. 
        However, the
        multimodeness of the light generated in these interferometers results 
        in the destruction of their phase sensitivity, requiring advanced
        interferometric configurations for multimode light.
        Moreover, in contrast to the single-mode case, 
        time-ordering effects play an important role for the high-gain 
        regime in the multimode scenario and
        must be taken into account for a correct estimation 
        of the phase sensitivity.
        In this work, we present a theoretical description of
        spatially multimode \suoo
        interferometers operating at low and high parametric gains. 
        Our approach is based on a step-by-step solution of a
        system of integro-differential
        equations for each nonlinear interaction region.
        We focus on interferometers with
        diffraction compensation, where focusing elements such as a parabolic
        mirror are used to compensate for the divergence of the light.
        We investigate plane-wave and Gaussian pumping and
        show that for any parametric gain,
        there exists a region of phases for which the
        phase sensitivity surpasses the standard shot-noise scaling
        and discuss the regimes where it approaches the Heisenberg scale.
        Finally, we arrive at insightful analytical expressions for the phase
        sensitivity that are valid for both low and high parametric
        gain and demonstrate how it depends on the number
        of spatial modes of the system.
    \end{abstract}

    \maketitle
    
    \section{INTRODUCTION}
        In recent years, nonlinear  \suoo interferometers have become
        an important subject in quantum
        metrology~\cite{Manceau2017,Chekhova:16,Frascella2019,PhysRevA.94.063840,PhysRevA.96.053863,Ferreri2022}
        as they can beat 
        the shot noise limit (SNL), which defines the best
        phase sensitivity that can be achieved
        using classical
        light~\cite{Manceau2017,Chekhova:16}. 
        Compared to linear interferometers, such as the traditional Mach-Zehnder
        interferometer operating with coherent light and
        bounded by SNL in estimating phase 
        sensitivity, they employ nonlinear processes, namely, 
        parametric down-conversion (PDC) or four-wave mixing (FWM)
        to create quantum states of light that allow for
        achieving phase sensitivities below the SNL~\cite{Manceau2017}.
        \par
\begin{figure}[btp]%
            \hspace*{0.1cm}%
            \def\svgwidth{\linewidth}%
            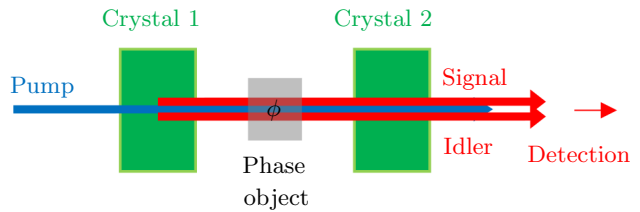%
            \caption{Simplified sketch of an \suoo interferometer consisting of two
            crystals. The PDC radiation (signal, idler) generated in the first
            crystal acquires some phase and is amplified/deamplified in the second
            crystal depending on the relative phase difference
            $\phi=\phi_p-\phi_s-\phi_i$ of the pump, signal and idler radiation.
            \label{fig:su11interferometerphase}}%
        \end{figure}

 A general sketch of the \suoo interferometer with two
        nonlinear crystals where the PDC process takes place
        is presented in \Cref{fig:su11interferometerphase}.
        Drawing an analogy with a linear
        interferometer, one can determine the SNL for a 
        nonlinear \suoo interferometer
        using the integral intensity of the light passing through the
        \textit{phase object} placed inside the
        interferometer~\cite{Chekhova:16,Frascella2019} 
        (the light generated by the first crystal):
        \begin{align}
            \Delta\phi_{\mathrm{SNL}} =
                \frac{1}{\sqrt{\langle \hat{N}_{\mathrm{tot}}^{\left(1\right)}\rangle}},
                \label{eq:def_snl}
        \shortintertext{where} 
            \langle \hat{N}_{\mathrm{tot}}^{\left(1\right)}\rangle 
                =\int\!\dd q\, \langle \hat{N}^{\left(1\right)} \of{q}\rangle
                \label{eq:def_Ntot_op}
        \end{align}
        is the total number of photons generated by the first crystal and
        $\langle \hat{N}^{\left(1\right)}\of{q}\rangle$ is the corresponding
        photon
        number distribution over the transverse momentum $q$.
        Throughout this paper, we will use the
        superscript~$^{\left(1\right)}$ to refer to
        quantities related to the first crystal and,
        analogously,~$^{\left(2\right)}$ for
        the second crystal.
        \par Using nonlinear interferometers, it is possible to surpass the
        shot-noise scaling defined by \Cref{eq:def_snl} and achieve the
        \textit{Heisenberg limit} or \textit{Heisenberg scaling} given
        by~\cite{Manceau2017,PhysRevA.94.063840,PhysRevA.33.4033,DemkowiczDobrzaski2012}
        \begin{align}\label{eq:heisenberg_scaling}
            \Delta\phi_\mathrm{H} &\propto \frac{1}{\langle \hat{N}_{s,\mathrm{tot}}^{\left(1\right)}\rangle},
        \end{align}
        which beats the shot-noise scaling for large
        intensities of light
        $\langle \hat{N}_{\mathrm{tot}}^{\left(1\right)}\rangle$~\cite{DemkowiczDobrzaski2012}.
        \par Most of the early studies focuses on the theoretical description of
        single-mode and two-mode
        interferometers~\cite{Manceau2017,PhysRevA.94.063840,Frascella2019}. 
        In the single-mode regime, the phase sensitivity of
        nonlinear interferometers
        can be improved by increasing the parametric gain of the
        nonlinear processes and by unbalancing their
        gains~\cite{Manceau2017,PhysRevA.96.053863}.
        However, in general, PDC and FWM couple many
        plane-wave modes and
        result in the generation of multimode
        light~\cite{Manceau2017,Christ_2014,Ferreri2021}, 
        which requires a 
        proper engineering of nonlinear interferometers 
        based on such
        multimode sources.
        Indeed, in the multimode case, to
        achieve a phase sensitivity below SNL, an
        appropriate dispersion
        compensation in the 
        frequency domain~\cite{Ferreri2021,Ferreri2022}
        or diffraction compensation in the
        spatial domain~\cite{Frascella2019} 
        must be performed. At the same
        time, in the multimode scenario, increasing
        the parametric gain brings 
        time-ordering effects
        into play~\cite{PRR2,Christ_2013} requiring a proper
        theoretical description of
        nonlinear interferometers at high gain.
        \par In this work, we theoretically study the phase
        sensitivity of spatially multimode
        \suoo interferometers at low and high parametric gains. Our 
        approach is based on the solution of
        the system of integro-differential equations for the
        plane-wave operators. We investigate
        various interferometric configurations and pump widths
        in order to show how the multimodeness of the light
        affects the phase sensitivity.
        \par For the multimode scenario discussed in this paper, we consider
        the phase sensitivity $\Delta\phi$ based on the output integral
        intensity of the interferometer. The phase sensitivity is
        usually defined via the error propagation relation
        as~\cite{gerry_knight_2004,Ferreri2021,Manceau2017}:
        \begin{align}\label{eq:def_delta_phi}
            \Delta\phi &= \frac{\Delta \hat{N}_{\mathrm{tot}}}{\left| \frac{\dd \langle \hat{N}_{\mathrm{tot}}\rangle}{\dd \phi} \right|} = \frac{\sqrt{\iint\!\dd q\,\dd q'\,\cov\of{q,q'}}}{\left| \frac{\dd \langle \hat{N}_{\mathrm{tot}}\rangle}{\dd \phi} \right|},
        \end{align}
        where 
        $\langle \hat{N}_{\mathrm{tot}}\rangle$,
        analogously to \Cref{eq:def_Ntot_op},
        is the output integral light intensity of the
        interferometer and the \textit{covariance} is given by
        \begin{align}\label{eq:def_cov}
            \cov\of{q,q'}= \langle \hat{N}(q)\hat{N}(q') \rangle - \langle \hat{N}(q)\rangle \langle \hat{N}(q') \rangle.
        \end{align}
        In the following discussion, we will consider the  
        phase sensitivity normalized with respect to the shot noise limit:
        \begin{align}
            f &= \frac{\Delta\phi}{\Delta\phi_{\mathrm{SNL}}}. \label{eq:def_f}
        \end{align}
        \par This paper is organized as follows.
        \Cref{sec:hg_su11} presents our theoretical approach for the description
        of high-gain \suoo interferometers based on the
        integro-differential equations
        and explains the concept of \textit{compensated} \suoo interferometers.
        Throughout this paper, we will assume that the signal and
        idler photons are distinguishable in some
        degree of freedom. However, as is shown in
        \Cref{sec:integ_cov_deg_case}, this assumption 
        does not affect the presented results.
        Additional details regarding the approach for the
        numerical solution of the 
        integro-differential equations are provided in 
        \Cref{sec:integ_diffeq_sol_details}.
        In \Cref{sec:pw_intro}, we first derive analytical
        solutions of the integro-differential equations in the limit of a
        plane-wave pump. Using these equations, the behavior of the optimal
        phase sensitivity is analyzed for both compensated and non-compensated
        \suoo interferometers. This discussion is then extended in
        \Cref{sec:finwidth}
        to a finite-width Gaussian pump. Here, we additionally obtain
        intriguing connections between the transfer functions describing
        the PDC process in each of both
        crystals and the Schmidt modes of such squeezers
        (see also \Cref{sec:analytic_impl_comp}).
        In \Cref{sec:comparison_pw_fw}, we then continue with
        a comparison of the optimal phase sensitivities of
        plane-wave and finite-width pumping with respect to the
        parametric gain, and
        investigate the width of the phase range for which the
        supersensitivity is achieved. Finally, we draw our
        conclusion in \Cref{sec:conclusion}.

    \section{HIGH-GAIN SU(1,1) INTERFEROMETER}\label{sec:hg_su11}
        In the Heisenberg representation, for both low and 
        high parametric gain, 
        the PDC process can be
        described by solving a set of coupled integro-differential
        equations for the signal/idler
        plane-wave operators $\hat{a}_s/\hat{a}_i$~\cite{PRR2}.
        In the case of a Gaussian pump
        $E_p\of{x,z,t}=E_0 \E^{-\frac{x^2}{2\sigma^2}} \E^{\I\left(k_p z-\omega_p t\right)}$,
        these equations are given by~\cite{PRR2}: 
        \begin{subequations}
            \label{eq:integ_diffeqs_both}
            \begin{gather}
                \begin{split}
                    \frac{\dd \hat{a}_s\of{q_s,L,\omega_s}}{\dd L} =\ & \Gamma \int\!\dd q_i\, 
                            \E^{-\frac{\left(q_s+q_i\right)^2\sigma^2}{2}} \\
                            &\times h\of{q_s,q_i,L} \hat{a}_i^{\dagger}\of{q_i,L,\omega_p-\omega_s},
                        \label{eq:integ_diffeqs_s}
                \end{split}\\
                \begin{split}
                    \frac{\dd \hat{a}_i^{\dagger}\of{q_i,L,\omega_p-\omega_s}}{\dd L} =\ & \Gamma \int\!\dd q_s\,
                        \E^{-\frac{\left(q_s+q_i\right)^2\sigma^2}{2}} \\
                        &\times h^{*}\of{q_s,q_i,L} \hat{a}_s\of{q_s,L,\omega_s},
                    \label{eq:integ_diffeqs_i}
                \end{split}
            \end{gather}
        \end{subequations}
        where $L$ is the integration variable, that is,
        the Cartesian coordinate axis in the collinear
        (longitudinal) direction parallel to the pump
        radiation.
        $\Gamma$ is the theoretical
        gain parameter proportional to the field amplitudes, 
        pump width~$\sigma$ and the nonlinear
        susceptibility; $q_{s/i}$ are the transverse wavevector
        s of the signal/idler
        photons and $h\of{q_s,q_i,L}$ is a
        function describing the phase matching of the PDC process.
        Note that the full width at half maximum (FWHM)
        of the intensity distribution of the Gaussian pump
        is given by $2\sqrt{\ln 2}\sigma$.
        \par The plane-wave operators obey the bosonic commutation relations:
        \begin{subequations}
            \begin{align}
                [\hat{a}_s\of{q_s,L,\omega_s}, \hat{a}_s^{\dagger}\of{q_s',L,\omega_s}] &= \delta\of{q_s-q_s'}, \label{eq:commrel_ss_finwidth} \\
                [\hat{a}_i\of{q_i,L,\omega_i}, \hat{a}_i^{\dagger}\of{q_i',L,\omega_i}] &= \delta\of{q_i-q_i'}, \label{eq:commrel_ii_finwidth} \\
                [\hat{a}_s\of{q_s,L,\omega_s}, \hat{a}_i^{\dagger}\of{q_i',L,\omega_i}] &= 0. \label{eq:commrel_si_finwidth}
            \end{align}
        \end{subequations}
        Note that the last commutation relation implies that the signal and idler
        photons are distinguishable with respect to some degree
        of freedom (frequency, polarization).
        More precisely, we will assume that the frequencies of the signal
        and idler photons are
        not identical but sufficiently close to each other so that their
        refractive indices 
        are coincide.
        \par With this assumption, the solution to this set of
        integro-differential equations
        has the form~\cite{PRR2}: 
        \begin{subequations}
            \begin{align}
                \begin{split}
                    \hat{a}_s\of{q_s,L,\omega_s} =\ & \hat{a}_s\of{q_s} + \int\!\dd q_s'\,\eta\of{q_s,q_s',L} \hat{a}_s\of{q_s'} \\
                    &+ \int\!\dd q_i'\,\beta\of{q_s,q_i',L} \hat{a}_i^{\dagger}\of{q_i'},
                    \label{eq:as_dagger_sol}
                \end{split}\\
                \begin{split}
                    \hat{a}_i^{\dagger}\of{q_i,L,\omega_i} =\ & \hat{a}_i^{\dagger}\of{q_i} + \int\!\dd q_i'\,
                        \eta^{*}\of{q_i,q_i',L} \hat{a}_i^{\dagger}\of{q_i'} \\
                        &+ \int\!\dd q_s'\,\beta^{*}\of{q_i,q_s',L} \hat{a}_s\of{q_s'},
                        \label{eq:ai_dagger_sol}
                \end{split}
            \end{align}
        \end{subequations}
        where $\hat{a}_s\of{q_s}$ and $\hat{a}_i\of{q_i}$ are the
        initial signal and idler
        plane-wave operators,
        while $\eta$ and $\beta$ are the complex-valued
        transfer functions depending on the transverse
        wavevectors and the crystal length.
        \par Note that this form of the solution with only two transfer functions
        $\eta$ and $\beta$ requires that the phase matching
        function is symmetric with respect to the wave vectors, namely,
        $h\of{q_s,q_i,L}=h\of{q_i,q_s,L}$, which is fulfilled
        for the type-I PDC process
        considered in this paper since, as we mentioned above, the refractive indices
        of the signal and idler photons are identical. The general solution
        to the system of integro-differential
        equations~\eqref{eq:integ_diffeqs_both}
        requires four transfer functions, so that the functions appearing in
        \Cref{eq:ai_dagger_sol} are no longer the complex conjugates of the functions
        appearing in \Cref{eq:as_dagger_sol}~\cite{Christ_2013,PRA102}.
        \par Plugging the solution in the form of
        \Cref{eq:as_dagger_sol,eq:ai_dagger_sol} into
        the integro-differential equations~\eqref{eq:integ_diffeqs_both}
        yields two equivalent\footnote{This is due to
        the fact that the input and the output operators are only connected
        via two distinct functions and their complex conjugates.}
        sets of coupled integro-differential equations for 
        $\eta$, $\beta$ and their complex conjugates,
        one of which reads
        \begin{subequations}
            \label{eq:integ_diffeqs_eta_beta_both}
            \begin{align}
                \begin{split}
                    \frac{\dd \beta\of{q_s,q_i',L}}{\dd L} &= \Gamma \int\!\dd q_i\,
                        \E^{-\frac{\left(q_s+q_i\right)^2\sigma^2}{2}} \\
                    &\qquad\times h\of{q_s,q_i,L} \tilde{\eta}^{*}\of{q_i,q_i',L},
                    \label{eq:integ_diffeqs_eb_beta}
                \end{split}\\
                \begin{split}
                    \frac{\dd \tilde{\eta}^{*}\of{q_i,q_i',L}}{\dd L} &= \Gamma \int\!\dd q_s\,
                        \E^{-\frac{\left(q_s+q_i\right)^2\sigma^2}{2}} \\
                    &\qquad\times h^{*}\of{q_s,q_i,L} \beta\of{q_s,q_i',L},
                    \label{eq:integ_diffeqs_eb_eta}
                \end{split}
            \end{align}
        \end{subequations}
        where 
        \begin{align}
            \tilde{\eta}^{*}\of{q_i,q_i',L}\eqdef\eta^{*}\of{q_i,q_i',L}+\delta\of{q_i-q_i'}.
        \end{align}
        This system of equations is easier to solve 
        numerically than the one given by
        \Cref{eq:integ_diffeqs_s,eq:integ_diffeqs_i}
        because it is no longer operator-valued. Note that this system is
        also similar to the ones already derived in
        Refs.~\cite{Christ_2013,PRA102} for the
        frequency domain.
        \par Using \Cref{eq:as_dagger_sol}, the mean photon
        number distribution (intensity distribution) of the signal photon can be
        expressed in terms of the transfer
        function $\beta$ ~\cite{PRR2}:
        \begin{align}
           \langle\hat{N}_s\of{q_s}\rangle &= \int\!\dd q_i' \left|\beta\of{q_s,q_i',L}\right|^2.
           \label{eq:intens_finwidth}
        \end{align}
        Similarly, the covariance is given by
        \begin{align}
            \begin{split}
                \cov\of{q_s,q_s'} &= \left|\int\!\dd q_i'\, 
                    \beta\of{q_s,q_i',L}\beta^{*}\of{q_s',q_i',L}\right|^2 \\
                &\qquad+\delta\of{q_s-q_s'}  \langle  \hat{N}_s\of{q_s} \rangle.
                \label{eq:cov_finwidth}
            \end{split}
        \end{align}
        Note that this expression does not contain a signal-idler cross-correlation
        term, since
        we assumed that the signal and idler photons are distinguishable, see
        \Cref{eq:commrel_si_finwidth}. The second term, proportional to the intensity distribution, can be
        identified as the shot noise term and follows from the Dirac-delta commutation relation of the operators due to the light energy quantization,
        while the argument of the modulus squared 
        corresponds to the field amplitude (first-order) correlation function
        $G^{\left(1\right)}\of{q_s,q_s'}$~\cite{PhysRevA.102.053725}.
        \par In the completely degenerate case, where the photons of the signal and idler beam are
        indistinguishable,  $\Gamma$  in
        the system of the integro-differential equations ~\eqref{eq:integ_diffeqs_both}  must
        be replaced with $2\Gamma$ due to the appearance
        of an additional term in the
        equations during the evaluation of the commutators in the Heisenberg equations (see
        the derivation in Ref.~\cite{PRR2}). However, as will be seen later, this replacement
        does not affect the experimentally relevant parametric gain $G$ defined
        in \Cref{sec:finwidth,sec:pw_intro}. Hence, the transfer functions $\tilde{\eta}$
        and $\beta$ will remain unchanged, given the same value of $G$.
        \par Following from that, as  shown in \Cref{sec:integ_cov_deg_case},
        the integral covariance is increased by a factor of~$2$ due to the appearance of the cross-correlation term. Additionally, due to the photon indistinguishability,
        the total intensity spectrum should  be considered instead of the signal beam intensity, which leads to an increase in the integrated intensity by a factor of~$2$:
         $\langle\hat{N}\of{q}\rangle
        =\langle\hat{N}_s\of{q}\rangle+\langle\hat{N}_i\of{q}\rangle
        =2 \langle\hat{N}_s\of{q}\rangle$. 
        Ultimately however, the normalized phase sensitivity $f$
        will remain unchanged because these additional factors of~$2$ cancel each other out, 
        see \Cref{eq:def_delta_phi,eq:def_f}.
        \par To describe the entire  \suoo interferometer, we solve
        the systems of integro-differential equations separately for
        the first and for the second crystals with functions
        $h^{\left(1\right)}\of{q_s,q_i,L}$ and $h^{\left(2\right)}\of{q_s,q_i,L}$, respectively,
        taking into account that the output operators of the first
        crystal are the input operators for the second crystal. In both cases
        the system is solved for the region $L \in \left[0,L_1\right]$, where 
        $L_1$ is the (single) crystal length. Further details are given
        in \Cref{sec:integ_diffeq_sol_details}.

        \subsection{Non-compensated interferometer}
\begin{figure*}[hbt]%
                \def\svgwidth{0.5\linewidth}%
                \subfloat[]{\label{fig:ncomsetup}%
                    %% Creator: Inkscape 1.2.2 (732a01da63, 2022-12-09), www.inkscape.org
%% PDF/EPS/PS + LaTeX output extension by Johan Engelen, 2010
%% Accompanies image file 'ncomsetup.eps' (pdf, eps, ps)
%%
%% To include the image in your LaTeX document, write
%%   \input{<filename>.pdf_tex}
%%  instead of
%%   \includegraphics{<filename>.pdf}
%% To scale the image, write
%%   \def\svgwidth{<desired width>}
%%   \input{<filename>.pdf_tex}
%%  instead of
%%   \includegraphics[width=<desired width>]{<filename>.pdf}
%%
%% Images with a different path to the parent latex file can
%% be accessed with the `import' package (which may need to be
%% installed) using
%%   \usepackage{import}
%% in the preamble, and then including the image with
%%   \import{<path to file>}{<filename>.pdf_tex}
%% Alternatively, one can specify
%%   \graphicspath{{<path to file>/}}
%% 
%% For more information, please see info/svg-inkscape on CTAN:
%%   http://tug.ctan.org/tex-archive/info/svg-inkscape
%%
\begingroup%
  \makeatletter%
  \providecommand\color[2][]{%
    \errmessage{(Inkscape) Color is used for the text in Inkscape, but the package 'color.sty' is not loaded}%
    \renewcommand\color[2][]{}%
  }%
  \providecommand\transparent[1]{%
    \errmessage{(Inkscape) Transparency is used (non-zero) for the text in Inkscape, but the package 'transparent.sty' is not loaded}%
    \renewcommand\transparent[1]{}%
  }%
  \providecommand\rotatebox[2]{#2}%
  \newcommand*\fsize{\dimexpr\f@size pt\relax}%
  \newcommand*\lineheight[1]{\fontsize{\fsize}{#1\fsize}\selectfont}%
  \ifx\svgwidth\undefined%
    \setlength{\unitlength}{239.47599792bp}%
    \ifx\svgscale\undefined%
      \relax%
    \else%
      \setlength{\unitlength}{\unitlength * \real{\svgscale}}%
    \fi%
  \else%
    \setlength{\unitlength}{\svgwidth}%
  \fi%
  \global\let\svgwidth\undefined%
  \global\let\svgscale\undefined%
  \makeatother%
  \begin{picture}(1,0.34712455)%
    \lineheight{1}%
    \setlength\tabcolsep{0pt}%
    \put(0,0){\includegraphics[width=\unitlength]{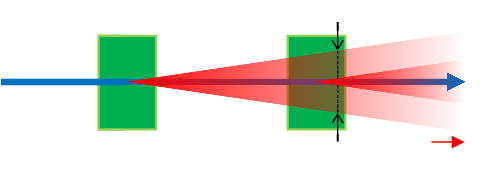}}%
    \put(0.55001029,0.31063888){\color[rgb]{0,0.69019608,0.31372549}\makebox(0,0)[lt]{\lineheight{1.25}\smash{\begin{tabular}[t]{l}Crystal 2\end{tabular}}}}%
    \put(0.17291843,0.31063888){\color[rgb]{0,0.69019608,0.31372549}\makebox(0,0)[lt]{\lineheight{1.25}\smash{\begin{tabular}[t]{l}Crystal 1\end{tabular}}}}%
    \put(0.39767766,0.08827838){\color[rgb]{1,0,0}\makebox(0,0)[lt]{\lineheight{1.25}\smash{\begin{tabular}[t]{l}PDC\end{tabular}}}}%
    \put(0.3670138,0.04130084){\color[rgb]{1,0,0}\makebox(0,0)[lt]{\lineheight{1.25}\smash{\begin{tabular}[t]{l}radiation\end{tabular}}}}%
    \put(-0.00386585,0.20728824){\color[rgb]{0,0.43921569,0.75294118}\makebox(0,0)[lt]{\lineheight{1.25}\smash{\begin{tabular}[t]{l}Pump\end{tabular}}}}%
    \put(0.79175683,0.00058696){\color[rgb]{1,0,0}\makebox(0,0)[lt]{\lineheight{1.25}\smash{\begin{tabular}[t]{l}Detection\end{tabular}}}}%
  \end{picture}%
\endgroup%
                }%
                \subfloat[]{\label{fig:compsetup}%
                    %% Creator: Inkscape 1.2.2 (732a01da63, 2022-12-09), www.inkscape.org
%% PDF/EPS/PS + LaTeX output extension by Johan Engelen, 2010
%% Accompanies image file 'compsetup.eps' (pdf, eps, ps)
%%
%% To include the image in your LaTeX document, write
%%   \input{<filename>.pdf_tex}
%%  instead of
%%   \includegraphics{<filename>.pdf}
%% To scale the image, write
%%   \def\svgwidth{<desired width>}
%%   \input{<filename>.pdf_tex}
%%  instead of
%%   \includegraphics[width=<desired width>]{<filename>.pdf}
%%
%% Images with a different path to the parent latex file can
%% be accessed with the `import' package (which may need to be
%% installed) using
%%   \usepackage{import}
%% in the preamble, and then including the image with
%%   \import{<path to file>}{<filename>.pdf_tex}
%% Alternatively, one can specify
%%   \graphicspath{{<path to file>/}}
%% 
%% For more information, please see info/svg-inkscape on CTAN:
%%   http://tug.ctan.org/tex-archive/info/svg-inkscape
%%
\begingroup%
  \makeatletter%
  \providecommand\color[2][]{%
    \errmessage{(Inkscape) Color is used for the text in Inkscape, but the package 'color.sty' is not loaded}%
    \renewcommand\color[2][]{}%
  }%
  \providecommand\transparent[1]{%
    \errmessage{(Inkscape) Transparency is used (non-zero) for the text in Inkscape, but the package 'transparent.sty' is not loaded}%
    \renewcommand\transparent[1]{}%
  }%
  \providecommand\rotatebox[2]{#2}%
  \newcommand*\fsize{\dimexpr\f@size pt\relax}%
  \newcommand*\lineheight[1]{\fontsize{\fsize}{#1\fsize}\selectfont}%
  \ifx\svgwidth\undefined%
    \setlength{\unitlength}{239.75067902bp}%
    \ifx\svgscale\undefined%
      \relax%
    \else%
      \setlength{\unitlength}{\unitlength * \real{\svgscale}}%
    \fi%
  \else%
    \setlength{\unitlength}{\svgwidth}%
  \fi%
  \global\let\svgwidth\undefined%
  \global\let\svgscale\undefined%
  \makeatother%
  \begin{picture}(1,0.39754364)%
    \lineheight{1}%
    \setlength\tabcolsep{0pt}%
    \put(0,0){\includegraphics[width=\unitlength]{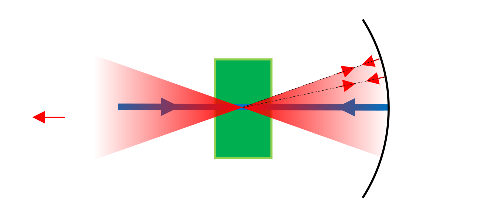}}%
    \put(0.42426984,0.31417602){\color[rgb]{0,0.69019608,0.31372549}\makebox(0,0)[lt]{\lineheight{1.25}\smash{\begin{tabular}[t]{l}crystal\end{tabular}}}}%
    \put(0.42882994,0.36109977){\color[rgb]{0,0.69019608,0.31372549}\makebox(0,0)[lt]{\lineheight{1.25}\smash{\begin{tabular}[t]{l}Single\end{tabular}}}}%
    \put(0.21721592,0.21157048){\color[rgb]{0,0.43921569,0.75294118}\makebox(0,0)[lt]{\lineheight{1.25}\smash{\begin{tabular}[t]{l}Pump\end{tabular}}}}%
    \put(0.79485576,0.18904603){\makebox(0,0)[lt]{\lineheight{1.25}\smash{\begin{tabular}[t]{l}Spherical\end{tabular}}}}%
    \put(0.82062506,0.14212229){\makebox(0,0)[lt]{\lineheight{1.25}\smash{\begin{tabular}[t]{l}mirror\end{tabular}}}}%
    \put(-0.00386142,0.18904603){\color[rgb]{1,0,0}\makebox(0,0)[lt]{\lineheight{1.25}\smash{\begin{tabular}[t]{l}Detection\end{tabular}}}}%
  \end{picture}%
\endgroup%
                }%
                \caption{(a) Non-compensated and (b) compensated
                    configurations of the \suoo interferometer. In the
                    non-compensated case, the radiation generated
                    in the first crystal diverges, which can be
                    considered as an internal loss mechanism
                    since the spatial overlap of the radiation generated
                    in two crystals is
                    reduced, as indicated by the black arrows.
                    Importantly, note that even if the distance
                    between the crystals
                    vanishes, such an overlap is still not perfect 
                    (the radiation also diverges inside the crystals).
                    In the compensated case, the spherical mirror
                    reflects the radiation back into the
                    crystal and compensates for the divergence.}%
            \end{figure*}
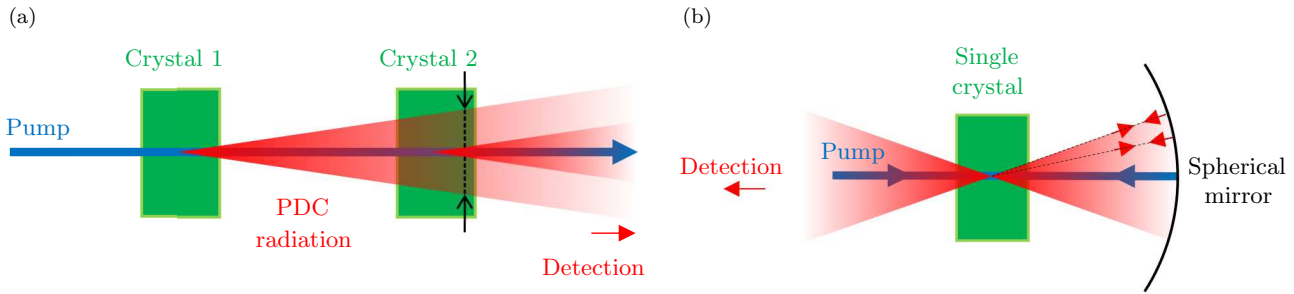

            An \suoo interferometer in its conventional form (without any focusing optical
            elements) is presented in \Cref{fig:ncomsetup}. It consists of two PDC sections,
            for example nonlinear crystals, separated by some spatial
            region in which the measured object is placed and
            the pump signal and idler radiation acquire some phase $\phi$. 
            This phase can be simply induced by an air gap between the two crystals. However, 
            more generally, electro-optical modulation or any kind
            of linear material between the
            crystals can be used to induce such a phase shift.
            \par Moreover, in a general description, the
            phase~$\phi\of{q_s,q_i}$ 
            might depend on the transverse wave vectors~$q_s$ and~$q_i$
            [for example, for large distances between the crystals in
            \Cref{fig:ncomsetup}].
            However, for simplicity, we assume that
            the distance between two crystals is small enough,
            so that we can restrict
            ourselves to phase shifts given by an effective
            constant phase~$\phi$. 
            This allows us to provide a direct comparison with the
            compensated interferometer and to obtain analytical
            expressions for the phase sensitivity in the
            finite-width pump case, see
            \Cref{sec:results_fw_comp} and \Cref{sec:integ_diffeq_sol_details}.
            \par In the multimode case, due to the divergence of the light,
            the radiation generated before (in the first crystal) and
            after (in the second crystal) the phase shift does not
            completely overlap, even if the distance between the
            crystals approaches zero, resulting in imperfect
            interference, which can be regarded
            as an internal loss mechanism.
            In the following, we will refer to such an
            interferometer as a \textit{non-compensated} \suoo interferometer.
            \par For the non-compensated interferometer, 
            the functions~$h$ that determine the dynamics
            of the field operators of the first and the second crystal are then given by
            \begin{subequations}
                \begin{align}
                    h^{\left(1\right)}\of{q_s,q_i,L} &= \E^{\I \Delta k\of{q_s,q_i} L}, \label{eq:h_sing_cryst} \\
                    h^{\left(2\right)}\of{q_s,q_i,L} &= \E^{\I \Delta k\of{q_s,q_i} \left[L+L_1\right]} \E^{\I \phi},
                    \label{eq:h_noncomp}
                \end{align}
            \end{subequations}
            respectively, where 
            $\Delta k\of{q_s,q_i}= \sqrt{k_p^2-\left(q_s+q_i\right)^2}- \sqrt{k_s^2-q_s^2} - \sqrt{k_i^2-q_i^2}$
            is the wavevector mismatch inside the PDC section
            and we suppose that the distance between the crystals is $d=0$.

        \subsection{Compensated interferometer}
            To compensate for this divergence, various focusing optical
            elements can be used, such as a spherical mirror as shown
            in \Cref{fig:compsetup} or the
            \textit{4f-optical system of lenses}~\cite{goodman2005}.
            Indeed, such focusing optical elements change the wavefront of
            light and add a quadratic phase depending on the transverse
            coordinate~$x$, namely, $\E^{-\I \frac{x^2}{2 F}}$, where~$F$
            is the focal length~\cite{goodman2005}.
            Therefore, if the crystal is placed
            at the~$2F$ position with respect to the spherical mirror (or two
            crystals placed at the~$F$ positions with respect to the 4f-optical
            system of lenses), the quadratic phase introduced by the mirror
            compensates for the quadratic phase of light, which
            leads to a change in the wavefront from convex to concave.
            In this case, taking
            into account the phase compensation and the wavefront
            modification, the $h$-function for the second
            crystal is given by
            \begin{align}
                h^{\left(2\right)}\of{q_s,q_i,L} &= \E^{-\I \Delta k\of{q_s,q_i} \left[L-L_1\right]} \E^{\I \phi}.
                     \label{eq:h_comp}
            \end{align}
            We define this kind of \suoo interferometer with 
            divergence compensation as a \textit{compensated} 
            \suoo interferometer.

    \section{PLANE-WAVE PUMP}\label{sec:pw_intro}
        We start our analysis by considering the
        plane-wave pump case, where analytical expressions for the output
        plane-wave operators can be obtained.
        Formally, the transition to the plane-wave pump case can be performed
        by taking the limit $\sigma\to\infty$ in 
        the system of integro-differential equations~\eqref{eq:integ_diffeqs_both}:
        \begin{align}
            \begin{split}
                \underbrace{\Gamma_0\frac{\sigma}{\sqrt{2\pi}}}_{=\Gamma}\E^{-\frac{\left(q_s+q_i\right)^2\sigma^2}{2}}
                    \xrightarrow{\sigma \to \infty} \Gamma_0\delta\of{q_s+q_i}.
            \end{split}
        \end{align}
        The resulting delta-function eliminates the
        integrals in the integro-differential
        equations~\eqref{eq:integ_diffeqs_both}
        and allows for an analytical treatment of the problem.
        \par More rigorously, the
        following set of coupled differential equations can be obtained
        by repeating the derivation presented in Ref.~\cite{PRR2} for
        a plane-wave pump:
        \begin{subequations}
            \label{eq:diffeqs_pw_0_both}
            \begin{align}
                \frac{\dd \hat{a}_s\of{q_s,L,\omega_s}}{\dd L} = \Gamma_0 h\of{q_s,L}
                    \hat{a}_i^{\dagger}\of{-q_s,L,\omega_p-\omega_s},
                    \label{eq:diffeq_pw_op_s}\\
                \frac{\dd \hat{a}_i^{\dagger}\of{-q_s,L,\omega_p-\omega_s}}{\dd L} = \Gamma_0
                    h^{*}\of{q_s,L} \hat{a}_s\of{q_s,L,\omega_s},
                    \label{eq:diffeq_pw_op_i}
            \end{align}
        \end{subequations}
        where $h\of{q_s,L}\eqdef h\of{q_s,-q_s,L}$. Note that in this case, each signal transverse
        momentum $q_s$ is connected with only 
        one fixed idler transverse momentum, $q_i=-q_s$.
        The commutation relations for the plane-wave operators
        are given by
        \Cref{eq:commrel_ss_finwidth,eq:commrel_ii_finwidth,eq:commrel_si_finwidth}.
        \par Similarly to the finite-width case presented in \Cref{sec:hg_su11},
        the solutions to the system of differential equations~\eqref{eq:diffeqs_pw_0_both} can be written
        in the
        form~\cite{PhysRevA.69.023802,Brambilla2001,RevModPhys.71.1539,PhysRevA.98.053827}: 
        \begin{subequations}
            \label{eq:sol_ops_pw_both}
            \begin{align}
                \begin{split}
                    \hat{a}_s\of{q_s,L,\omega_s} &= \tilde{\eta}_{\mathrm{pw}}\of{q_s,L}
                        \hat{a}_s\of{q_s} \\
                        &\qquad+ \beta_{\mathrm{pw}}\of{q_s,L}
                        \hat{a}_i^{\dagger}\of{-q_s},
                        \label{eq:sol_ops_pw_s}
                \end{split}\\
                \begin{split}
                    \hat{a}_i^{\dagger}\of{-q_s,L,\omega_i} &= \tilde{\eta}^{*}_{\mathrm{pw}}\of{-q_s,L}
                        \hat{a}_i^{\dagger}\of{-q_s} \\
                        &\qquad+ \beta^{*}_{\mathrm{pw}}\of{-q_s,L} \hat{a}_s\of{q_s}.
                        \label{eq:sol_ops_pw_i}
                \end{split}
            \end{align}
        \end{subequations}
        Again, this form of the solution requires 
        $h\of{q_s,-q_s}=h\of{-q_s,q_s}$, or, more specifically,
        $h\of{q_s}=h\of{-q_s}$
        and leads to two equivalent
        sets of differential equations for the
        $\beta_{\mathrm{pw}}$ and $\tilde{\eta}^{*}_{\mathrm{pw}}$ functions,
        one of which reads:
        \begin{subequations}
            \label{eq:diffeqs_eta_beta_2_pw_both}
            \begin{align}
                \frac{\dd \beta_{\mathrm{pw}}\of{q_s,L}}{\dd L} &= 
                    \Gamma_0 h\of{q_s,L} \tilde{\eta}^{*}_{\mathrm{pw}}\of{-q_s,L},
                    \label{eq:diffeqs_eta_beta_2_pw_s}
                    \\
                \frac{\dd \tilde{\eta}^{*}_{\mathrm{pw}}\of{-q_s,L}}{\dd L} &= 
                    \Gamma_0 h^{*}\of{q_s,L} \beta_{\mathrm{pw}}\of{q_s,L}.
                    \label{eq:diffeqs_eta_beta_2_pw_i}
            \end{align}
        \end{subequations}
        \par Using \Cref{eq:sol_ops_pw_s}, the distribution of the mean number of signal photons is given by
        \begin{align}
           \langle\hat{N}_s\of{q_s}\rangle &= \left|\beta_{\mathrm{pw}}\of{q_s,L}\right|^2 \delta\of{0}.\label{eq:intens_pw_delta}
        \end{align}
        The divergent factor $\delta\of{0}$ appears due to the Dirac-delta
        commutation relations for the plane-wave operators and the infinite
        transverse size of the system~\cite{PhysRevA.98.053827}. 
        Below, to avoid this divergence, we instead consider the density of quantities, namely,
        the intensity and covariance per transverse length of the system, which are well-defined even in the limit of
         $L_x\to\infty$.
        \par For example, the density of the intensity spectrum of the signal radiation is given by
        \begin{align}
            \NN_s\of{q_s} &= \left|\beta_{\mathrm{pw}}\of{q_s,L}\right|^2. \label{eq:intens_pw_finite}
        \end{align}
        \par Similarly, the covariance density of the signal radiation can be expressed via the intensity as
        (see also Ref.~\cite{klyshko1988photons}):
        \begin{subequations}
            \begin{align}
                \cov_\mathrm{pw}\of{q_s,q_s'} = \CC\of{q_s} \delta\of{q_s-q_s'},
                \label{eq:cov_pw_cc}
            \shortintertext{where}
                \CC\of{q_s} = \NN_s\of{q_s} \left[ 1 + \NN_s\of{q_s}\right].
                    \label{eq:pw_cov_general_finite_spectral}
            \end{align}
        \end{subequations}
        A more detailed discussion of this divergence problem
        and the derivation of
        \Cref{eq:intens_pw_finite,eq:cov_pw_cc,eq:pw_cov_general_finite_spectral} is shown in \Cref{sec:divergence_treatment}.
        \par Solving the coupled differential equations~\eqref{eq:diffeqs_eta_beta_2_pw_both} 
        for a single crystal
        with \Cref{eq:h_sing_cryst} and the initial conditions
        $\tilde{\eta}_{\mathrm{pw}}\of{q_s}=1$ and $\beta_{\mathrm{pw}}\of{q_s}=0$
        yields the transfer functions connecting the input and output operators~\cite{klyshko1988photons}:
        \begin{widetext}
            \begin{subequations}\label{eq:pw_solutions_sing_cryst__singular}
                \begin{align}
                    \beta_{\mathrm{pw}}^{\left(1\right)}\of{q_s,L} &= \frac{2\Gamma_0}{g\of{q_s}}
                        \sinh\of{\frac{L_1g\of{q_s}}{2}}\E^{\I \frac{\Delta k\of{q_s}L_1}{2}}, \\
                    \tilde{\eta}_{\mathrm{pw}}^{\left(1\right)}\of{q_s,L} &=
                        \left[\cosh\of{\frac{L_1g\of{q_s}}{2}}-\frac{\I\Delta
                        k\of{q_s}}{g\of{q_s}}\sinh\of{\frac{L_1g\of{q_s}}{2}}\right]
                        \E^{\I \frac{\Delta k\of{q_s}L_1}{2}},
                \end{align}
            \end{subequations}
        \end{widetext}
        where
        \begin{align}
            g\of{q_s} &= \sqrt{4 \Gamma_0^2-\Delta k^2\of{q_s}}, \label{eq:def_g}
        \end{align}
        and $\Delta k\of{q_s}\eqdef\Delta k\of{q_s,-q_s}$.
        Note that $\Delta k\of{q_s}=\Delta k\of{-q_s}$.
        \par Then, the density of the intensity distribution of the signal
        beam after the first crystal is given by:
        \begin{align}
            \NN_s^{\left(1\right)}\of{q_s} &=
                \left[\frac{2\Gamma_0}{g\of{q_s}}\sinh\of{\frac{L_1 g\of{q_s}}{2}}\right]^2.
            \label{eq:intens1c}
        \end{align}
        Note that although $g\of{q_s}$ can take both real
        and purely imaginary values, the intensity distribution is always real and
        positive due to the $\sinh$-term.
        Therefore, it is not necessary to write $\left|\,\cdot\,\right|^2$ on the right
        hand side of \Cref{eq:intens1c}.
        Note that \Cref{eq:pw_solutions_sing_cryst__singular,eq:def_g,eq:intens1c}
        derived above coincide with the results already found in a similar fashion
        in Refs.~\cite{PhysRevA.69.023802,Brambilla2001,RevModPhys.71.1539,klyshko1988photons}.
        \par To analyze the effect of the focusing element
        (spherical mirror) on the output spectra and the
        phase sensitivity, we first start by extending the
        plane-wave-pump analytical treatment to the \suoo
        interferometer without any compensation elements
        and then compare the non-compensated scheme with
        its compensated counterpart.
        \par In this work, we consider an \suoo interferometer consisting of
        BBO crystals of length $L_1=\SI{2}{\milli\meter}$,
        pumped by a laser with a wavelength of $\SI{354.6}{\nano\meter}$.
        To obtain a
        connection between the theoretical gain parameter $\Gamma_0$ and the
        experimental gain, the collinear output intensity 
        $\NN_s^{\left(1\right)}\of{0}$  of a single
        crystal is fitted by the function 
        $y\of{\Gamma_0}=B\sinh^2\of{A\Gamma_0}$. The experimental gain $G$
        is then given by $G=A\Gamma_0$~\cite{PRR2}. For plane-wave
        pumping and perfect phase matching in the collinear direction [that is, $\Delta k\of{0}=0$],
        it is immediately clear from \Cref{eq:intens1c} that $A=L_1$ and $B=1$.
        \par If instead of $\Gamma_0$, the prefactor in
        the system of coupled differential
        equations~\eqref{eq:diffeqs_eta_beta_2_pw_both} 
        is $c\Gamma_0\eqdef\Gamma_0'$, where $c>0$, 
        the fitting constant would still be given by $A=L_1$ for a 
        fit with respect to $\Gamma_0'$ and the experimental gain would be defined
        as $G=A\Gamma_0'$. The fitting constant $B$ would be given by $c^2$. Therefore,
        the definition of the experimental gain is independent on the
        scaling factor $c$. For the degenerate case
        mentioned in \Cref{sec:hg_su11},  the scaling factor is $c=2$. Therefore, all results presented in this
        section would be the same for the degenerate case where the signal and
        idler photons are indistinguishable.

        \subsection{Non-compensated scheme}
            In order to find a solution in the case of the
            non-compensated \suoo interferometer, we analytically solve
            the system of differential
            equations~\eqref{eq:diffeqs_pw_0_both} using
            \Cref{eq:h_sing_cryst,eq:h_noncomp},
            and obtain the output plane-wave operators at the
            output of the \suoo interferometer.
            In this case, the form of the solution
            solution~\eqref{eq:sol_ops_pw_both} has the following
            transfer functions:
            \begin{widetext}
                \begin{subequations}
                    \begin{align}
                        \begin{split}
                            \beta_{\mathrm{pw}}\of{q_s} &= -\frac{4\Gamma_0}{g^2\of{q_s}}
                                \left[\Delta k\of{q_s} \sin\of{\frac{\phi}{2}}
                                \sinh\of{\frac{L_1g\of{q_s}}{2}}
                                - g\of{q_s} \cos\of{\frac{\phi}{2}}
                                \cosh\of{\frac{L_1g\of{q_s}}{2}}\right] \\
                            &\qquad\times\sinh\of{\frac{L_1g\of{q_s}}{2}} \E^{\I\left(L_1\Delta
                                k\of{q_s}+\frac{\phi}{2}\right)},
                                \label{eq:beta_eta_pw_beta}
                        \end{split}\\
                        \begin{split}
                            \tilde{\eta}_{\mathrm{pw}}\of{q_s} &= -\frac{\E^{\I (\Delta k\of{q_s} L_1+\phi )}}{g^2} 
                           \bigl\{ -2 \Gamma_0^2 \left[\cosh\of{L_1 g\of{q_s}} -1\right] \\
                                &\qquad+\E^{-\I \phi} \left[-2 \Gamma_0^2+\left(2 \Gamma_0^2-g^2\right)
                              \cosh\of{L_1 g\of{q_s}}+ \I \Delta k\of{q_s} g\of{q_s} \sinh\of{L_1 g\of{q_s}}\right]\bigr\}.
                                \label{eq:beta_eta_pw_eta}
                        \end{split}
                    \end{align}
                \end{subequations}
                These functions allow us to calculate the density of the output intensity distribution of 
                the signal beam:
                \begin{align}\label{eq:intens_out_su11_ncomp}
                    \begin{split}
                        \NN_s\of{q_s} &= \NN_s^{\left(1\right)}\of{q_s}
                        \pmb{\bigg\lvert}\E^{\I \phi} \left[\cosh\of{\frac{L_1 g\of{q_s}}{2}} + 
                        \frac{\I \Delta k\of{q_s}}{g \of{q_s}}\sinh\of{\frac{L_1 g\of{q_s}}{2}}\right] \\
                        &\qquad+\left[\cosh\of{\frac{L_1 g\of{q_s}}{2}} - 
                        \frac{\I \Delta k\of{q_s}}{g \of{q_s}}\sinh\of{\frac{L_1 g\of{q_s}}{2}}\right]{\pmb{\bigg\rvert}}^2.
                    \end{split}
                \end{align}
            \end{widetext}
            \par The covariance density for the signal beam can
            be found by plugging \Cref{eq:intens_out_su11_ncomp}
            into \Cref{eq:pw_cov_general_finite_spectral}
            with the use of ~\Cref{eq:cov_pw_cc}.
            \par
\begin{figure*}
                \subfloat{\label{fig:pw_intens_ncom_0}%
                    \includegraphics[width=0.5\linewidth]
                        {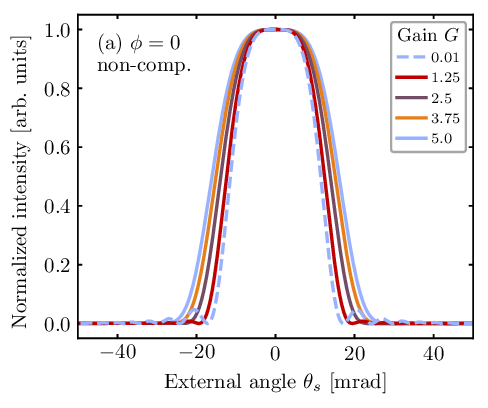}}%
                \subfloat{\label{fig:pw_intens_ncom_05}%
                    \includegraphics[width=0.5\linewidth]
                        {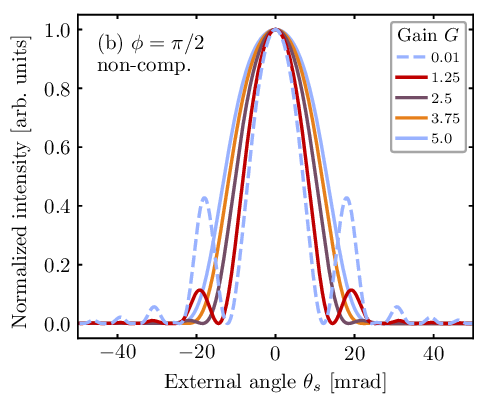}}\\
                \subfloat{\label{fig:pw_intens_ncom_1}%
                    \includegraphics[width=0.5\linewidth]
                        {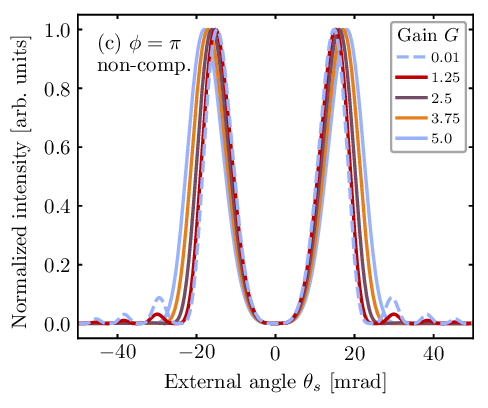}}%
                \subfloat{\label{fig:pw_intens_comp}%
                    \includegraphics[width=0.5\linewidth]
                        {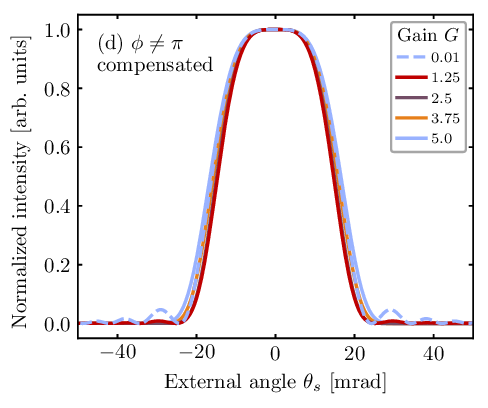}}%
                \caption{Calculated intensity distributions for different phases 
                    between the crystals and gains in the case of  plane-wave pumping. 
                    (a)--(c)~Non-compensated interferometer
                    and (d)~compensated interferometer.
                    In the compensated case, the intensity profile is only scaled
                    as~${\sim\cos^2\of{\phi/2}}$, see \Cref{eq:intens_pw_su11_comp_N,eq:intens_pw_su11_comp_xi}.
                    \label{fig:pw_intens}}%
            \end{figure*}

 \Cref{fig:pw_intens_ncom_0,fig:pw_intens_ncom_05,fig:pw_intens_ncom_1}
            present the intensity distributions for different gains and phases
            between the crystals according to \Cref{eq:intens_out_su11_ncomp}. They
            are plotted over the external angle $\theta_s$, which, in the case of
            small angles, is
            connected to the transverse wavevector via
            ${\theta_s\approx q_s/k_s^{\mathrm{vac}}}$, 
            where $k_s^{\mathrm{vac}}$ is the wavevector of the signal photons
            in vacuum. Note that throughout this work, we will use the external angles
            instead of the transverse wave vectors for all relevant plots.
            \par
\begin{figure*}[tb]
                \subfloat{\label{fig:pw_f_ncom}%
                    \includegraphics[width=0.495\linewidth]{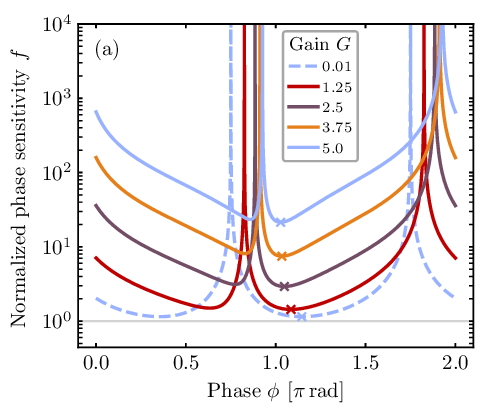}}%
                \subfloat{\label{fig:pw_f_comp}%
                    \includegraphics[width=0.5\linewidth]{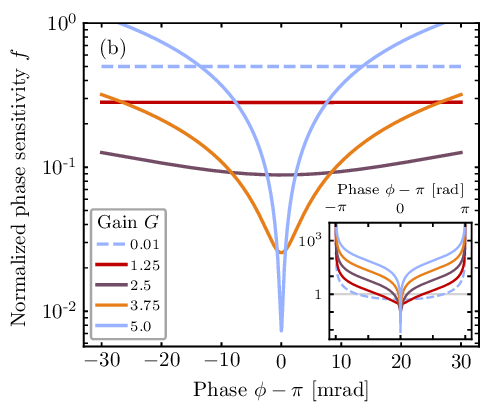}}%
                \caption{(a)~Calculated normalized phase sensitivity $f$ for
                    different gains for the non-compensated setup. For each gain, the crosses mark
                    the points where $f$ is minimized.
                    The horizontal thin gray line indicates the shot noise level ($f=1$).
                    The divergences occur at points where
                    the derivative of the integral intensity (density)
                    vanishes: ${\frac{\dd \NN_{s,\mathrm{tot}}}{\dd\phi}=0}$ [see
                    \Cref{eq:def_delta_phi,eq:integral_phot_dens_pw}].
                    (b)~Calculated normalized phase sensitivity $f$ for
                    different gains for the compensated setup. Note that the entire plotted
                    region is below the shot noise level. The inset shows the full phase
                    interval $\left[0,2\pi\right]$. As mentioned in the text, the minimal value of
                    $f$ is always reached at $\phi=\pi$. Furthermore, $f$ diverges for
                    ${\phi\to 2\pi k}, {k\in\mathbb{Z}}$, see \Crefrange{eq:Delta_phi_pw_comp_DP}{eq:Delta_phi_pw_comp_BB}.
                    In both cases, the pump is given by a plane wave.}%
            \end{figure*}

 One can observe that the intensity profiles broaden as
            the gain increases. The phase between the crystals strongly modifies
            the intensity profiles, leading to destructive interference
            for a certain range of angles. This is due to the fact that
            the radiation generated in each crystal has a quadratic phase
            with respect to the angle, see
            \Cref{eq:beta_eta_pw_beta,eq:beta_eta_pw_eta}. Even if
            the distance between the crystals is zero, such a quadratic
            phase is present and leads to a modification of the intensity
            distribution when various additional constant phases are applied.  
            The distributions of the covariance have a similar behavior
            to the intensity profiles and are shown in \Cref{sec:cov_plots_appendix_pw}.
            The normalized phase sensitivity is
            presented in \Cref{fig:pw_f_ncom}. It can be
            seen that the sensitivity is destroyed as the gain increases. 
            This is directly related to the fact that the
            quadratic phase of the radiation
            entering the second crystal is not compensated,
            that effectively acts as
            internal losses and destroys the phase sensitivity.
            With increasing gain,
            such losses become more pronounced.
            A similar behavior of the phase sensitivity 
            of the frequency multimode
            \suoo interferometer was observed in Refs.~\cite{Ferreri2021,Ferreri2022}.

        \subsection{Compensated scheme}\label{sec:results_pw_comp}
            In the case of the compensated \suoo interferometer, we analytically solve
            the system of differential
            equations~\eqref{eq:diffeqs_pw_0_both} with the use of \Cref{eq:h_sing_cryst}
            and \Cref{eq:h_comp}, and find the following transfer functions which connect 
            the output and the input plane-wave operators of the entire interferometer:
            \begin{widetext}
                \begin{subequations}
                    \begin{align}
                        \beta_{\mathrm{pw}}\of{q_s} &= \frac{\Gamma_0}{g \of{q_s}}
                            \left\{ \frac{\I\Delta k\of{q_s}}{g\of{q_s}} \left[ \cosh\of{L_1 g\of{q_s}} - 1 
                                    \right] + \sinh\of{L_1 g\of{q_s}} \right\}\left(1+\E^{\I\phi}\right), \\
                        \tilde{\eta}_{\mathrm{pw}}\of{q_s} &= 1+\frac{2\Gamma_0^2}{g^2\of{q_s}}
                            \left[\cosh\of{L_1 g\of{q_s}}-1\right]\left(1+\E^{\I \phi}\right).
                    \end{align}
                \end{subequations}
            \end{widetext}
            One can observe that in this case, both the functions $ \beta_{\mathrm{pw}}\of{q_s}$
            and $\tilde{\eta}_{\mathrm{pw}}\of{q_s}$ have a simple dependence on the phase $\phi$,
            which allows the output intensity to be written in a much more compact and simple 
            form compared to \Cref{eq:intens_out_su11_ncomp}:
            \begin{subequations}
                \begin{align}
                   \NN_s\of{q_s} = 4 \xi_{\mathrm{pw}}\of{q_s} \cos^2\of{\frac{\phi}{2}},
                   \label{eq:intens_pw_su11_comp_N}
                \shortintertext{where}
                    \xi_{\mathrm{pw}}\of{q_s} = \NN_s^{\left(1\right)}\of{q_s}
                        \left[1 + \NN_s^{\left(1\right)}\of{q_s}\right].
                    \label{eq:intens_pw_su11_comp_xi}
                \end{align}
            \end{subequations}
            Note that the entire intensity spectrum now scales as~${\sim\cos^2\of{\phi/2}}$ and is
            therefore identically zero for
            $\phi=\pi$, which implies perfect destructive interference for all angles
            of emission.
            \par The normalized intensity distribution for different gains is shown in
            \Cref{fig:pw_intens_comp}. For the chosen interval of gains,
            the width of the distribution remains almost unchanged.
            However, since the distribution of a single crystal broadens
            with increasing gain, then, according to \Cref{eq:intens_pw_su11_comp_N,eq:intens_pw_su11_comp_xi},
            the intensity distribution of the entire interferometer will also be
            broadened for higher values of gain.
            \par Substituting the expression for the intensity distribution given by 
            \Cref{eq:intens_pw_su11_comp_N,eq:intens_pw_su11_comp_xi} into
            \Cref{eq:cov_pw_cc,eq:pw_cov_general_finite_spectral}, 
            one can calculate
            the covariance in the compensated case. 
            The plots of the covariances for
            different interferometric phases and gains 
            are presented in \Cref{sec:cov_plots_appendix_pw} and show
            a similar behaviour to the intensity profiles.
            \par Finally, by replacing the intensity and the covariance in
            \Cref{eq:def_delta_phi} with their densities given in 
            \Cref{eq:intens_pw_su11_comp_N,eq:intens_pw_su11_comp_xi} 
            and \Cref{eq:cov_pw_cc,eq:pw_cov_general_finite_spectral}, respectively, it 
            is possible to obtain an analytical expression for the
            phase sensitivity density:
            \begin{subequations}
                \begin{align}
                    \Delta\Phi = \frac{\sqrt{
                        \AA_\mathrm{pw} + 4\BB_\mathrm{pw}\cos^2\of{\frac{\phi}{2}} }
                    }{2\AA_\mathrm{pw} \left|\sin\of{\frac{\phi}{2}}\right|},
                    \label{eq:Delta_phi_pw_comp_DP}
                \shortintertext{where}
                    \AA_\mathrm{pw} = \int\!\dd q_s\, \xi_{\mathrm{pw}}\of{q_s}
                        \label{eq:integ_intens_su11_pw} \\
                \intertext{is the density of the integral 
                intensity of the \suoo interferometer and }
                    \BB_\mathrm{pw} = \int\!\dd q_s \left[\xi_{\mathrm{pw}}\of{q_s}\right]^2.
                    \label{eq:Delta_phi_pw_comp_BB}
                \end{align}
            \end{subequations}
            \par Note that in the plane-wave pump case, the phase sensitivity
            defined via the integral quantities [\Cref{eq:def_delta_phi}]
            can be obtained by dividing the phase sensitivity density
            [\Cref{eq:Delta_phi_pw_comp_DP}]
            by~$\sqrt{L_x}$,
            which means that the integral phase sensitivity is
            length-dependent. However, the SNL-normalized phase
            sensitivity~$f$, given by \Cref{eq:def_f}, no longer
            depends on the considered transverse size of the system
            and coincides with the normalized phase sensitivity density.
            \par The SNL-normalized phase sensitivity~$f$ 
            is presented in \Cref{fig:pw_f_comp}.
            In this case, due to the full compensation, all spatial
            modes are deamplified
            simultaneously,  therefore, the
            phase sensitivity behaves similarly to the
            single-mode case~\cite{Manceau2017} and beats the shot noise limit. 
            Moreover, from \Cref{fig:pw_f_comp} it becomes apparent that the
            width~$\Delta$ of the phase range for which
            the SNL is overcome (${f < 1}$) gradually narrows as the parametric
            gain increases.
            Furthermore, this region is defined for all gains, which means that for all
            gain values (even for low gains), there exists some range of phases where the phase sensitivity beats the SNL,
            which is similar to early studies of the single-mode
            interferometer~\cite{PhysRevA.33.4033}.  
            \par From \Crefrange{eq:Delta_phi_pw_comp_DP}{eq:Delta_phi_pw_comp_BB} it can be easily seen that
            the best phase sensitivity (minimal values of~${\Delta\phi}$ and~$f$) is achieved
            for ${\phi=\pi}$, that is, at the dark fringe of the interferometer. Similarly,
            the optimal working point
            for more elaborate models
            (including losses) was found to be
            near the dark fringe~\cite{Manceau2017,PhysRevA.86.023844,PhysRevA.94.063840}.
            The value of the minimized SNL-normalized phase sensitivity  is then given by:
            \begin{subequations}
                \begin{align}
                    f_{\mathrm{pw,min}} &= \frac{1}{2}\sqrt{
                        \frac{\NN_{s,\mathrm{tot}}^{\left(1\right)}
                        }{\NN_{s,\mathrm{tot}}^{\left(1\right)}
                        + \NN_{s,2}^{\left(1\right)}}},
                        \label{eq:fmin_plane_wave_f}
                \end{align}
                where
                \begin{align}
                    \NN_{s,\mathrm{tot}}^{\left(1\right)} &= 
                        \int\!\dd q_s\, \NN_{s}^{\left(1\right)}\of{q_s},
                        \label{eq:fmin_plane_wave_Ntot}\\
                    \NN_{s,2}^{\left(1\right)} &= \int\!\dd q_s
                        \left[\NN_{s}^{\left(1\right)}\of{q_s}\right]^2.
                        \label{eq:fmin_plane_wave_N2}
                \end{align}
            \end{subequations}
            \par Notably, the expressions derived above only depend on the intensity spectra of the
            first crystal. This is due to the fact that both crystals have the same parameters
            (length, gain etc.). At the same time,  the perfect compensation  induces a symmetry,
            so that the parameters of the first crystal fully describe the system.
            \par The behavior of the optimal phase sensitivity as a function of the
            parametric gain~$G$ is discussed in \Cref{sec:comparison_pw_fw}
            where  a comparison to the finite-width pump case 
            and the Heisenberg scaling of the phase sensitivity is also drawn.

    \section{FINITE-WIDTH GAUSSIAN PUMP}\label{sec:finwidth}
        \par In this section, we extend our studies by considering a pump beam
        with a Gaussian profile with a FWHM of the
        intensity distribution of $\SI{50}{\micro\meter}$. 
        All other parameters are the same as
        in the plane-wave case, see \Cref{sec:pw_intro}.
        For a finite-width pump, the output signal and idler plane-wave
        operators can only be found by numerical integration of the system of
        integro-differential equations~\eqref{eq:integ_diffeqs_eta_beta_both}, as 
        described in \Cref{sec:integ_diffeq_sol_details}. From there, the 
        intensity distributions, covariances and the normalized phase
        sensitivities are calculated numerically.
        \par Analogously to the plane-wave case, the collinear
        intensity\footnote{Note that the factor~$\dd q$ has to be taken
        into account because ${\langle\hat{N}_s^{\left(1\right)}\of{0}\rangle}$
        corresponds to the photon \textit{density} evaluated at ${q=0}$.
        Multiplying by $\dd q$ gives the actual photon count.}
        ${\langle\hat{N}_s^{\left(1\right)}\of{0}\rangle\,\dd q}$
        from a single crystal must be fitted in order to obtain
        a connection between the theoretical gain parameter~$\Gamma$
        and the experimental gain~$G$. However, in the multimode
        regime with finite-width pumping, this procedure is not
        straightforward as for the plane-wave pump case, especially
        if a large range of gain values should be considered.
        To resolve this, instead of a single fit, many fits
        are performed, capturing different ranges of the
        parametric gain. More details regarding this procedure
        and exact values of~$A$ for the results presented
        in the following sections are given in
        \Cref{sec:A_G_dependence}.

        \par Similarly to the plane-wave pump (see \Cref{sec:pw_intro}), 
        the definition of the
        experimental gain is independent of any additional
        scaling factors $c$ in the system of integro-differential
        equations~\eqref{eq:integ_diffeqs_both}, since the theoretical
        gain constant can be
        redefined to absorb such factors. The remaining integro-differential
        equations are identical to the ones without the scaling factor $c$, so
         making a fit with respect to the newly defined theoretical gain
        constant $\Gamma'\eqdef c\Gamma$
        yields the same $A$. The definition of $G$ is then given by with
        respect to the scaled gain constant $G=A\Gamma'$, and therefore remains
        unchanged.

        \subsection{Non-compensated scheme}
\begin{figure*}
                \subfloat{\label{fig:fw_intens_ncom_0}%
                    \includegraphics[width=0.5\linewidth]
                        {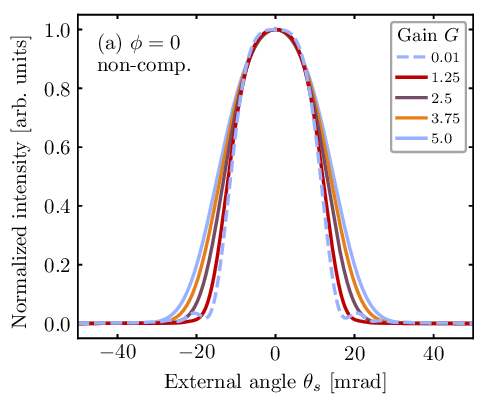}}%
                \subfloat{\label{fig:fw_intens_ncom_05}%
                    \includegraphics[width=0.5\linewidth]
                        {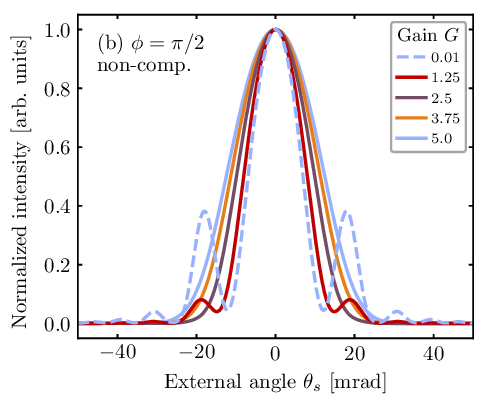}}\\
                \subfloat{\label{fig:fw_intens_ncom_1}%
                    \includegraphics[width=0.5\linewidth]
                        {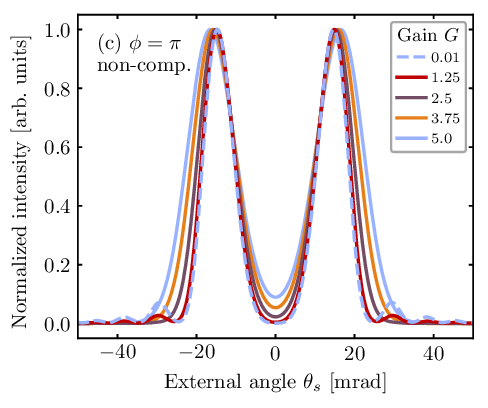}}%
                \subfloat{\label{fig:fw_intens_comp}%
                    \includegraphics[width=0.5\linewidth]
                        {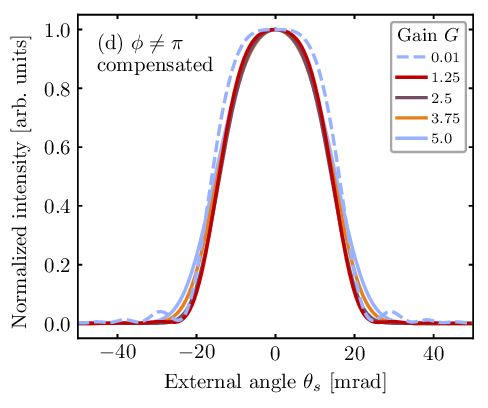}}%
                \caption{Calculated intensity distributions for a finite-width
                Gaussian pump with FWHM of $\SI{50}{\micro\meter}$ for
                the (a)--(c)~non-compensated interferometer
                and for the (d)~compensated interferometer for different phases.
                Same as for the plane-wave pump case, varying the phase, the intensity profiles
                of the compensated \suoo interferometer are
                only scaled as~$\sim\cos^2\of{\phi/2}$,
                see \Cref{eq:intens_su11_fw_N,eq:xi_def_beta_eta}, therefore, only one plot for all
                $\phi\neq\pi$ is presented.\label{fig:fw_intens}}%
            \end{figure*}

\begin{figure*}[tb]
                \subfloat{\label{fig:fw50_f_ncom}%
                    \includegraphics[width=0.495\linewidth]{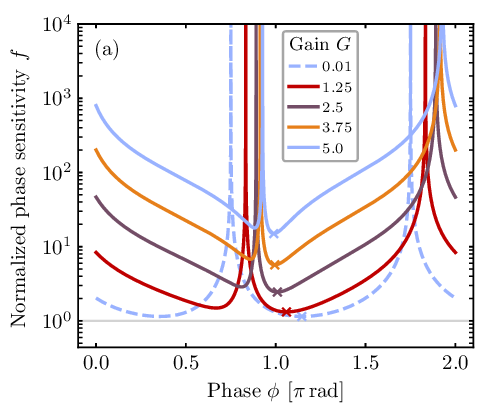}}%
                \subfloat{\label{fig:fw50_f_comp}%
                    \includegraphics[width=0.5\linewidth]{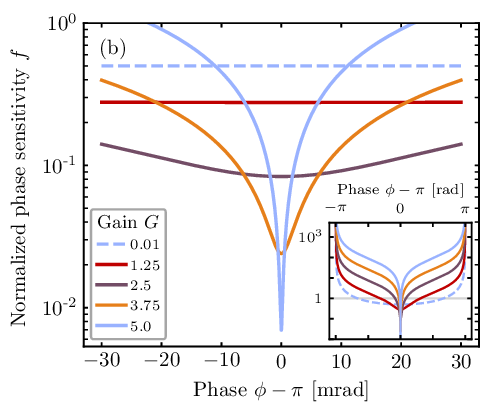}}%
                \caption{(a)~Calculated normalized phase sensitivity $f$ for
                    different gains for the non-compensated setup. For each gain, the crosses mark
                    the points where $f$ is minimized.
                    The horizontal thin gray line indicates the shot noise level ($f=1$).
                    The divergences occur at points where
                    ${\frac{\dd \langle\hat{N}_{s,\mathrm{tot}}\rangle}{\dd\phi}=0}$
                    [see \Cref{eq:def_delta_phi}].
                    (b)~Calculated normalized phase sensitivity $f$ for
                    different gains for the compensated setup. Note that the entire plotted
                    region is below the shot noise level. The inset shows the full phase
                    interval $\left[0,2\pi\right]$. As mentioned in the text, the minimal value of
                    $f$ is always reached at $\phi=\pi$. Furthermore, $f$ diverges for
                    ${\phi\to 2\pi k}, {k\in\mathbb{Z}}$, see \Cref{eq:analytic_expr_f_fw}.
                    In both cases, the pump is given by a Gaussian pump with an intensity-FWHM
                    of \SI{50}{\micro\meter}.}%
            \end{figure*}

            In the case of the non-compensated scheme, the profiles 
            of the intensity distribution for different 
            phases and gains are presented in
            \Cref{fig:fw_intens_ncom_0,fig:fw_intens_ncom_05,fig:fw_intens_ncom_1}.
            One can observe that the Gaussian pump profile modifies the intensity
            at the zero phase, bringing it
            closer to the Gaussian shape. Overall, for different phases,
            the graphs show a similar behavior to
            the plane-wave pump case. 
            However, since the number of modes is smaller in the case
            of a finite-width pump compared to the case
            of a plane-wave pump, the intensity distribution in the case
            of  finite width pumping broadens more slowly.
            The plots of the covariance are presented and discussed in
            \Cref{sec:cov_plots_appendix_fw}. \Cref{fig:fw50_f_ncom}
            presents profiles of the SNL-normalized phase sensitivity
            for different gains where no supersensitivity regions are observed.

        \subsection{Compensated scheme}\label{sec:results_fw_comp}
            For the compensated configuration,
            similarly to the plane-wave pump, it is possible to obtain
            simplified expressions for the intensity, the covariance and the phase
            sensitivity. To do this, relationships between the transfer functions
            of the first and the second crystal  are required.
            In the case of the perfect
            compensation, these relations are derived in
            \Cref{sec:analytic_impl_comp} and are given by:
            \begin{subequations}\label{eq:connection_eta_beta_comp_c1c2}
                \begin{align}
                    \tilde{\eta}^{\left(2\right)}_{\phi}\of{q,q'} &=
                        \left[\tilde{\eta}^{\left(1\right)}\of{q',q}\right]^{*}, \\
                    \beta^{\left(2\right)}_{\phi}\of{q,q'} &=
                        \E^{\I\phi}\beta^{\left(1\right)}\of{q',q}.
                \end{align}
            \end{subequations}
            To be precise, in the above expression, the transfer functions with the
            index~$^{\left(1\right)}$ (index~$^{\left(2\right)}$)
            connect the input and the output plane-wave operators of the
            first (second) crystal according
            to \Cref{eq:as_dagger_sol,eq:ai_dagger_sol}. The index $_\phi$ indicates
            the dependence on the interferometer phase
            for the second crystal\footnote{Technically, as it is shown
            in \Cref{sec:analytic_impl_comp}, the
            transfer function~$\tilde{\eta}^{\left(2\right)}$
            is independent of the interferometer phase. Nevertheless,
            we add this index to emphasize the phase dependence
            for the second crystal.}.
            \par In \Cref{sec:integ_diffeq_sol_details}, we derive expressions
            for the composite transfer functions of the entire
            interferometer ($\tilde{\eta}^{\left(\mathrm{SU}\right)}$ and $\beta^{\left(\mathrm{SU}\right)}$) in terms of the transfer functions
            of both separate crystals, see \Cref{eq:composite_tf_eta_nophase,eq:composite_tf_beta_nophase},
            which
            can be used to obtain a simplified expression for the
            intensity spectrum of the 
            entire interferometer via \Cref{eq:intens_finwidth}:
            \begin{subequations}
                \begin{align}
                    \langle\hat{N}_s\of{q_s}\rangle &= 4\cos^2\of{\frac{\phi}{2}} \int\!\dd q'\left|\xi\of{q_s,q'}\right|^2,
                    \label{eq:intens_su11_fw_N}
                \end{align}
                where
                \begin{align}
                    \xi\of{q_s,q'} &= \int\!\dd\bar{q}\,\beta^{\left(1\right)}\of{\bar{q},q_s}
                        \left[\tilde{\eta}^{\left(1\right)}\of{\bar{q},q'}\right]^{*}. \label{eq:xi_def_beta_eta}
                \end{align}
            \end{subequations}
            Importantly, note that the function $\xi\of{q_s,q'}$ no longer
            depends on $\phi$. This implies that,
            as in the plane-wave pump case, all spatial
            modes are amplified or deamplified 
            simultaneously, and  the intensity distribution is
            identically zero for $\phi=\pi$.
            Furthermore, the expression for the output intensity
            depends only  on the transfer functions
            of the first crystal.
            This is due to the  fact that the full compensation induces 
            a symmetry in the system, see \Cref{sec:analytic_impl_comp}.
            \par Plots of the intensity distributions for several parametric gain
            values can be found in \Cref{fig:fw_intens_comp}. For the chosen gain parameters, 
            the width of the intensity distribution remains almost unchanged, however, as the
            gain increases, its shape becomes more Gaussian. This is associated with a decrease
            in the number of modes with increasing gain.
            \par Further simplifications   are possible by considering the Schmidt decomposition
            of the transfer functions. As it was shown in Refs.~\cite{Christ_2013,PRA102},
            there exists a joint Schmidt decomposition (Bloch-Messiah reduction)
            for the transfer functions $\beta$ and $\tilde{\eta}$, which reads
            \begin{subequations}
                \begin{align}
                    \beta\of{q,q'} &= \sum_{n} \sqrt{\Lambda_{n}} u_{n}\of{q}\psi_{n}\of{q'},
                        \label{eq:sdecomp_beta} \\
                    \tilde{\eta}\of{q,q'} &= \sum_{n} \sqrt{\tilde{\Lambda}_{n}}
                        u_{n}\of{q}\psi_{n}^{*}\of{q'},
                        \label{eq:sdecomp_eta}
                \end{align}
            \end{subequations}
            where $\sqrt{\Lambda_{n}}$ and $\sqrt{\tilde{\Lambda}_{n}}$ are the singular values
            of $\beta$ and $\tilde{\eta}$, respectively, $u_n\of{q}$ 
            are the functions associated with the output Schmidt operators (we  denote these functions as the output Schmidt modes)
            and $\psi_n\of{q'}$ are associated with the input Schmidt operators  (the input Schmidt modes) of the considered system.
            See also \Cref{sec:integ_cov_deg_case_fw} for further details.
            \Cref{eq:sdecomp_eta,eq:sdecomp_beta} extend the results for the Schmidt decomposition 
            already found for the frequency domain in Refs.~\cite{Christ_2013,PRA102} to the
            spatial domain. Note that the eigenvalues $\Lambda_{n}$ and $\tilde{\Lambda}_{n}$
            are connected via $\tilde{\Lambda}_{n}=1+\Lambda_{n}$~\cite{Christ_2013}.
            \par Applying \Cref{eq:sdecomp_eta,eq:sdecomp_beta} to \Cref{eq:intens_su11_fw_N,eq:xi_def_beta_eta},
            the total intensity of the
            \suoo interferometer can be written as 
            \begin{subequations}
                \begin{align}
                    \langle\hat{N}_{s,\mathrm{tot}}\rangle = 4\AA\cos^2\of{\frac{\phi}{2}},
                \shortintertext{where}
                    \AA = \sum_n \Lambda_n^{\left(1\right)} \left(1+\Lambda_n^{\left(1\right)}\right). \label{eq:def_AA}
                \end{align}
            \end{subequations}
            \par Similarly, the covariance can be written as
            \begin{align}\label{eq:cov_fw_phase_separated}
                \begin{split}
                    \cov\of{q_s,q_s'} &= 16\cos^4\of{\frac{\phi}{2}}
                        \int\!\dd \bar{q}\left|\xi\of{q_s,\bar{q}}
                        \xi^{*}\of{q_s',\bar{q}}\right|^2 \\
                        &\qquad+ \delta\of{q_s-q_s'} \langle\hat{N}_s\of{q_s}\rangle.
                \end{split}
            \end{align}
            Plots of the covariance for different gains and phases are presented and discussed
            in \Cref{sec:cov_plots_appendix_fw}.
            \par The integral covariance can be written in the following form:
            \begin{subequations}
                \begin{align}
                    \begin{split}
                        \iint\!\dd q_s\,\dd q_s'\cov\of{q_s,q_s'}
                        = 4\cos^2\of{\frac{\phi}{2}}\!
                            \left[\AA+4\BB\cos^2\of{\frac{\phi}{2}}\right],
                    \end{split}
                \end{align}
                where
                \begin{align}
                    \BB = \sum_n \left[\Lambda_n^{\left(1\right)} \left(1+\Lambda_n^{\left(1\right)}\right)\right]^2.
                \end{align}
            \end{subequations}
            \par Finally, combining the formulas for the integral
            intensity and covariance  presented in this section, 
            it is possible to derive
            an analytic expression for the phase sensitivity in 
            the case of perfect compensation and finite-width pumping:
            \begin{align}\label{eq:analytic_expr_f_fw}
                \Delta\phi &= \frac{\sqrt{\AA+4\BB\cos^2\of{\frac{\phi}{2}}}}
                        {2\AA\left|\sin\of{\frac{\phi}{2}}\right|}.
            \end{align}
            Note that the expression above has a similar form to the
            phase sensitivity in terms of the
            density of quantities presented in 
            \Crefrange{eq:Delta_phi_pw_comp_DP}{eq:Delta_phi_pw_comp_BB} for the plane-wave case.
            Profiles of the SNL-normalized phase sensitivity~$f$ for different gains
            are shown in \Cref{fig:fw50_f_comp}.
            One can observe that as the gain increases,
            the phase supersensitivity range width~$\Delta$ 
            (the range of phases for which 
            the phase sensitivity beats the shot noise limit)
            gradually decreases. However, for all presented gains, there
            exists some region where the phase sensitivity
            beats the shot noise level.
            \par From \Cref{eq:analytic_expr_f_fw} it can be seen
            that, similarly to the plane-wave
            pump case, the best (optimal) phase sensitivity is
            achieved for the optimal point ${\phi=\pi}$, 
            that is, at the dark fringe of the interferometer (see also
            Refs.~\cite{Manceau2017,PhysRevA.86.023844,PhysRevA.94.063840}).
            For this phase, the SNL-normalized phase sensitivity is given by
            \begin{align}\label{eq:fmin_fw}
                f_{\mathrm{min}} &= \frac{1}{2}
                    \sqrt{\frac{\langle\hat{N}_{s,\mathrm{tot}}^{\left(1\right)}\rangle}{\AA}}.
            \end{align}
            A more detailed analysis of the behavior of~$f_{\mathrm{min}}$
            and~$\Delta$ as a function of~$G$
            is given in the next \Cref{sec:comparison_pw_fw}, where we also
            draw a comparison to the
            plane-wave pump case and the Heisenberg scaling of the
            phase sensitivity.
    
    \section{COMPARISON OF THE OPTIMAL PHASE SENSITIVITIES} \label{sec:comparison_pw_fw}
        In this section, to compare the plane-wave and finite-width pumping regimes,  we only  consider the compensated
        setup, because the non-compensated setup shows no
        supersensitivity for any phase, as seen in the previous sections.
        \begin{figure}[tb]
            \includegraphics[width=\linewidth]{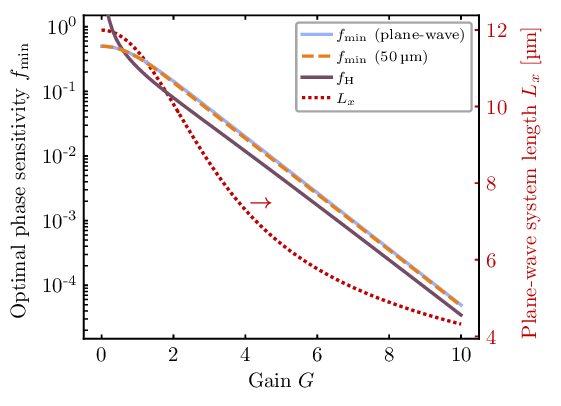}%
            \caption{Optimal phase sensitivities~$f_{\mathrm{min}}$ of the compensated
                \suoo interferometer for the plane-wave pump and the finite-width
                Gaussian pump with a FWHM of the intensity 
                distribution of \SI{50}{\micro\meter}
                versus the parametric gain~$G$. $f_\mathrm{H}$~corresponds
                to the expected asymptotic Heisenberg scaling normalized with respect
                to the SNL, see \Cref{eq:heisenberg_scaling_fH}.
                $L_x$~(right-hand vertical axis) is the transverse 
                length of the system pumped by
                the plane-wave pump to reach the same output intensity after the first crystal
                as in the case of the system pumped by the finite-width pump.
                Note that  due to the assumed equality of intensities, the Heisenberg scalings coincide in both cases. See also
                \Cref{sec:divergence_treatment} for more details on~$L_x$.
                \label{fig:fopt_plot}}%
        \end{figure}%
        \par As it was discussed in \Cref{sec:pw_intro}, in the plane-wave pump case,
        the integral quantities depend on the transverse length $L_x$ of the system.
        Therefore, to ensure a proper comparison of the two cases, we assume that the plane-wave pump system has the required transverse length $L_x$, so that its integral number of photons after the first crystal is equal to the integral number of photons in the  case of finite-width pumping for the same value of the parametric gain.
        The dependence of~$L_x$ on the gain is presented in \Cref{fig:fopt_plot}.
        For more details see \Cref{sec:divergence_treatment}.
        \par The assumed equality of the light intensities after the first
        crystal leads to the same Heisenberg limit for both cases which,
        in accordance with \Cref{eq:heisenberg_scaling}, we define as
        \begin{align}\label{eq:heisenberg_scaling_Deltaphi}
            \Delta\phi_\mathrm{H} &= \frac{1}{2 \langle \hat{N}_{s,\mathrm{tot}}^{\left(1\right)}\rangle} = \frac{1}{2L_x\NN^{\left(1\right)}_{s,\mathrm{tot}}}.
        \end{align}
        Thus, the SNL-normalized Heisenberg scaling is given by
        \begin{align}\label{eq:heisenberg_scaling_fH}
            f_\mathrm{H} &= \frac{1}{2 \sqrt{\langle \hat{N}_{s,\mathrm{tot}}^{\left(1\right)}\rangle}} = \frac{1}{2\sqrt{L_x \NN^{\left(1\right)}_{s,\mathrm{tot}}}}.
        \end{align}
        \par To compare the optimal phase sensitivity in the plane-wave
        and the finite-width
        cases, we show~$f_{\mathrm{min}}$ for both cases over a range of parametric
        gains in \Cref{fig:fopt_plot}. One can observe that, as
        the parametric gain increases,
        the optimal phase sensitivity approaches the Heisenberg scaling
        for both pumps. However, the phase sensitivity for the finite width
        pump slightly surpasses the phase sensitivity for the plane-wave pump.
        The reason for this is the different number of modes in two cases, namely, the number of modes in the finite-width case is smaller compared to
        the plane-wave case.
        \par To see this more clearly, we define the Schmidt number, which is
        a measure of the effective number of modes, for the first
        crystal in the finite-width case
        as~\cite{PRA91,PhysRevX.10.031063,PhysRevA.102.053725}:
        \begin{subequations}
            \begin{align}
                K^{\left(1\right)} = \left[\sum_n\left(\lambda_n^{\left(1\right)}\right)^2\right]^{-1},
                \label{eq:schmidt_number_fw_K}
            \end{align}
            where
            \begin{align}
                \lambda_n^{\left(1\right)} = \frac{\Lambda_n^{\left(1\right)}}{\sum_k\Lambda_k^{\left(1\right)}}
                \label{eq:schmidt_number_fw_lambda}
            \end{align}
        \end{subequations}
        are the normalized eigenvalues obtained from the decomposition of the transfer
        function~$\beta^{(1)}$ of the first crystal, see \Cref{eq:sdecomp_beta}.
        We can then rewrite \Cref{eq:def_AA} as
        \begin{align}
            \AA &= \left(1+\frac{\langle\hat{N}_{s,\mathrm{tot}}^{\left(1\right)}\rangle}{K^{\left(1\right)}}\right)
                \langle\hat{N}_{s,\mathrm{tot}}^{\left(1\right)}\rangle, \label{eq:def_AA_K}
        \end{align}
        since ${\langle\hat{N}_{s,\mathrm{tot}}^{\left(1\right)}\rangle
        =\sum_k\Lambda_k^{\left(1\right)}}$.
        This allows us to express the optimal phase sensitivity via the
        Schmidt number as
        \begin{align}\label{eq:fmin_pw}
            f_{\mathrm{min}} = \frac{1}{2\sqrt{
                1+\frac{\langle\hat{N}_{s,\mathrm{tot}}^{\left(1\right)}\rangle}{K^{\left(1\right)}} }}.
        \end{align}
        From this expression it becomes clear that the phase sensitivity is optimized
        for high intensities (high parametric gains) and small effective mode numbers.
        In the extreme case of a single mode (${K=1}$) with an eigenvalue
        $\Lambda_0^{\left(1\right)}$, the optimal phase sensitivity is
        given by
        \begin{align}
            f_{\mathrm{min}} = \frac{1}{2\sqrt{1+\Lambda_0^{\left(1\right)}}},
        \end{align}
        which approaches the Heisenberg limit~$f_\mathrm{H}$ for ${\Lambda_0^{\left(1\right)}\gg 1}$.
        \par In the plane-wave case, due to the strong correlations
       between the signal and idler photons, it is no longer possible
        to define Schmidt modes as in the finite-width pump case.
        However, by analogy, we find from
        \Crefrange{eq:fmin_plane_wave_f}{eq:fmin_plane_wave_N2}:
        
        \begin{subequations}
            \begin{align}
                f_{\mathrm{pw,min}} = 
                    \frac{1}{2\sqrt{1+\frac{L_x\NN_{s,\mathrm{tot}}^{\left(1\right)}}
                        {K_\mathrm{pw}^{\left(1\right)}}}},
                    \label{eq:schmidt_number_plane_wave_f}
            \shortintertext{where}
                K_\mathrm{pw}^{\left(1\right)} = L_x\frac{\left[\int\!\dd q_s\, \NN_{s}^{\left(1\right)}\of{q_s}\right]^2}{\int\!\dd q_s
                        \left[\NN_{s}^{\left(1\right)}\of{q_s}\right]^2}
                \label{eq:schmidt_number_plane_wave_K}
            \end{align}
        \end{subequations}
        can be defined as the Schmidt number in the plane-wave pump case.
        Alternatively, we can write it as
        \begin{subequations}
            \begin{align}
                K_\mathrm{pw}^{\left(1\right)} = L_x\KK\ofb{\beta^{\left(1\right)}\of{q_s}},
            \shortintertext{with the following functional defined for some function~$u$: }
                \KK\ofb{u\of{q_s}} = \frac{\left[\int\!\dd q_s\, \left|u\of{q_s}\right|^2\right]^2}{\int\!\dd q_s\left|u\of{q_s}\right|^4}.
            \end{align}
        \end{subequations}
       This provides us with an expression for
        $K_\mathrm{pw}^{\left(1\right)}$ which has already
        been found and discussed in Ref.~\cite{Horoshko2012} for the low-gain
        regime.
        \par
\begin{figure}[tb]
            \includegraphics[width=\linewidth]{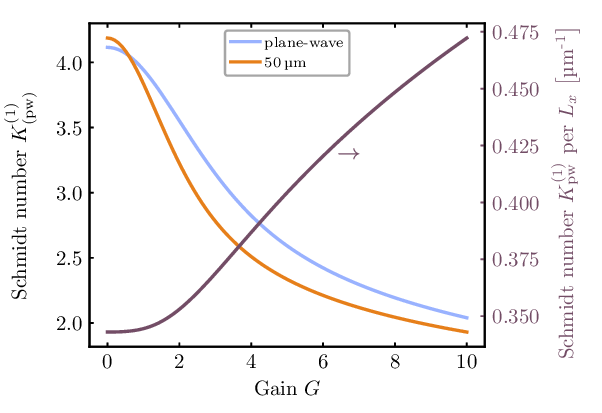}%
            \caption{The Schmidt number for the plane-wave pump case
                [from \Cref{eq:schmidt_number_plane_wave_f,eq:schmidt_number_plane_wave_K}] and finite-width
                pump case [from
                \Cref{eq:schmidt_number_fw_K,eq:schmidt_number_fw_lambda}]
                for
                several parametric gains. Additionally, on the right-hand
                vertical axis, the plane-wave Schmidt number per length~$L_x$
                ($K^{\left(1\right)}_{\mathrm{pw}}/L_x$) is presented.
                The required values of~$L_x$ are shown in \Cref{fig:fopt_plot}.
                \label{fig:snum_plots}}%
        \end{figure}

 Plots of the Schmidt numbers for plane-wave and finite-width
        pumping are shown in
        \Cref{fig:snum_plots}. As expected, 
        the Schmidt number
        gradually decreases as the parametric gain increases~\cite{PRR2}.
        For $G>1$, the Schmidt number for the finite-width pump is lower
        than for the plane-wave pump. Therefore,
        assuming equal intensities after the first crystal in
        both cases, the phase sensitivity
        for finite-width pumping surpasses the phase sensitivity for
        plane-wave pumping as it is shown in \Cref{fig:fopt_plot}. 
        \par
\begin{figure}[tb]
            \includegraphics[width=\linewidth]{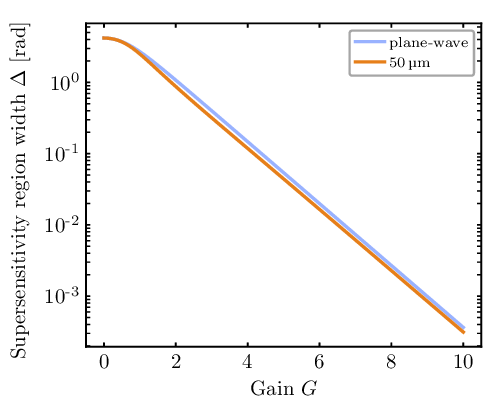}%
            \caption{The phase supersensitivity range width~$\Delta$
                (length of the phase interval where~${f < 1}$) versus the
                parametric gain for the plane-wave and finite-width
                Gaussian pump.\label{fig:Delta_plot}}%
        \end{figure}

 The phase supersensitivity range width~$\Delta$ as the
        function of parametric gain~$G$ is shown in \Cref{fig:Delta_plot}
        for both finite-width pumping and plane-wave pumping.
        Clearly, at low gains the width for both cases coincides.
        As the gain increases, plane-wave pumping leads to a slightly
        narrower phase sensitivity region. However, in both
        cases,~$\Delta$ drops quickly and decreases by about one order
        of magnitude with each increase of the gain~$G$ by two.

    \section{CONCLUSION}\label{sec:conclusion}
        In this work, we present a theoretical description of the high-gain
        multimode \suoo interferometer, which, 
        however, is also valid for the low-gain regime. Our approach is
        based on the solution
        of the system of integro-differential equations for the plane-wave
        operators and allows us to introduce a set of gain-dependent Schmidt
        modes. We investigate the
        spatial properties and phase sensitivity of multimode nonlinear
        interferometers in different configurations (with
        and without diffraction compensation) for various pump widths.
        \par In the case of a plane-wave pump, the system of
        integro-differential equations is reduced to the set of ordinary
        differential equations with an analytical solution. At finite-width
        pumping, the system of integro-differential equations is evaluated
        numerically to find the corresponding transfer functions. In addition,
        the use of the derived relations between the transfer functions of
        individual crystals and the entire interferometer, allows us to
        obtain compact expressions for the phase sensitivity for
        finite-width pumping. Finally, the minimized phase sensitivity
        for the plane-wave and the finite-width pump is compared for
        different gains. 
        \par We demonstrate that for the non-compensated interferometers,
        the phase sensitivity worsens as the parametric gain increases and never
        surpasses the shot noise scaling, which is due to incomplete overlapping
        of the radiation of the first and the second crystals. However, for the
        compensated interferometers, both finite-width 
        and plane-wave pumping result in a phase sensitivity that always
        exceeds SNL, so that
        the optimal phase sensitivity approaches the Heisenberg scaling for
        large parametric gains.
        However, the phase width
        of the supersensitivity region becomes increasingly narrow as the gain increases.
        To counteract this narrowing, as well as to optimize
        the phase sensitivity in the case of optical and detection losses,
        unbalanced \suoo interferometers  \cite{Manceau2017} can be considered, which may be
        the subject of future research.

    \section*{ACKNOWLEDGMENTS}
        We acknowledge financial support of the
        Deutsche Forschungsgemeinschaft (DFG) via 
        project SH 1228/3-1 and via the TRR 142/3, project C10.
        We also thank the PC\textsuperscript{2} (Paderborn
        Center for Parallel Computing) for
        providing computation time.

    \appendix

    \section{Integral covariance including the cross-correlation}\label{sec:integ_cov_deg_case}
        \subsection{Finite-width pump}\label{sec:integ_cov_deg_case_fw}
            As was described in \Cref{sec:hg_su11}, it is assumed throughout this paper
            that the signal and idler photons are distinguishable in some degree of freedom,
            which is then reflected in the fact that the signal and idler plane-wave
            operators commute, see \Cref{eq:commrel_si_finwidth}.
            In the fully degenerate regime, this commutator is instead given by
            \begin{align}
                [\hat{a}_s\of{q_s,L,\omega_s}, \hat{a}_i^{\dagger}\of{q_i',L,\omega_i}] &=
                    \delta\of{q_s-q_i'}. \label{eq:commrel_si_deg_finwidth}
            \end{align}
            This does not influence the intensity distributions for the signal and idler beams
            as \Cref{eq:intens_finwidth} remains unchanged.
            However, the non-zero commutator between the signal and idler operators 
            will lead to an additional cross-correlation term in the covariance,
            \Cref{eq:cov_finwidth}. The full expressions then reads
            \begin{subequations}
                \begin{align}
                    \cov\of{q,q'} = \auto\of{q,q'} + \cross\of{q,q'},
                \end{align}%
                where%
                \begin{align}
                    \begin{split}
                        \auto\of{q,q'} &= \left|\int\!\dd \bar{q}\,
                            \beta\of{q,\bar{q},L}\beta^{*}\of{q',\bar{q},L}\right|^2 \\
                                &\qquad+\delta\of{q-q'}  \langle  \hat{N}\of{q} \rangle,
                            \label{eq:fw_auto_corr}
                    \end{split} \\
                    \cross\of{q,q'} &= \left|\int\!\dd \bar{q}\,
                        \beta\of{q,\bar{q},L}\tilde{\eta}\of{q',\bar{q},L}\right|^2.
                        \label{eq:fw_cross_corr}
                \end{align}
            \end{subequations}
            Here, we have dropped the signal and idler labels since the photons are
            indistinguishable.
            Note that the auto-correlation term is the expression already appearing in
            \Cref{eq:cov_finwidth}, while the cross-correlation term is new.
            \par To further evaluate the integral covariance, it is necessary to
            consider the joint Schmidt decomposition of the transfer
            functions $\beta$ and $\tilde{\eta}$
            which was introduced in
            \Cref{eq:sdecomp_eta,eq:sdecomp_beta}. The existence 
            of such a joint decomposition was demonstrated
            in Refs.~\cite{Christ_2013,PRA102}.
            Utilizing the joint Schmidt decomposition and the
            commutation relations for the plane-wave operators, useful
            relations between the transfer functions
            can be derived:
            \begin{subequations}
                \begin{align}
                    \begin{split}
                        &\int\!\dd \bar{q}\, \beta\of{q,\bar{q},L}\beta^{*}\of{q',\bar{q},L} + \delta\of{q-q'}\\
                            &\qquad=\int\!\dd \bar{q}\, \tilde{\eta}\of{q,\bar{q},L}\tilde{\eta}^{*}\of{q',\bar{q},L}
                            \label{eq:eta_beta_rels_from_comm_bbee}
                    \end{split}
                \end{align}
                for \Cref{eq:commrel_ss_finwidth,eq:commrel_ii_finwidth} and
                \begin{align}
                    \int\!\dd \bar{q}\, \tilde{\eta}\of{q,\bar{q},L}\beta\of{q',\bar{q},L}
                        = \int\!\dd \bar{q}\, \beta\of{q,\bar{q},L}\tilde{\eta}\of{q',\bar{q},L}
                            \label{eq:eta_beta_rels_from_comm_ebeb}
                \end{align}
            \end{subequations}
            for the commutation relation
            \begin{align}
                [\hat{a}_s\of{q_s,L,\omega_s}, \hat{a}_i\of{q_i,L,\omega_i}] &= 0. \label{eq:commrel_si_finwidth_nodagger}
            \end{align}
            These relations have already
            been used to obtain more compact expressions for the covariance in
            \Cref{eq:cov_finwidth,eq:fw_auto_corr,eq:fw_cross_corr}.
            Similar equations for the frequency domain have been found
            in Ref.~\cite{Christ_2013}.
            Furthermore, note that there are two additional relationships which can
            be derived by applying the commutation constraints to the inverse
            transform of \Cref{eq:as_dagger_sol,eq:ai_dagger_sol}, see \Cref{sec:analytic_impl_comp}.
            \par Applying the joint Schmidt
            decomposition from \Cref{eq:sdecomp_eta,eq:sdecomp_beta} to \Cref{eq:fw_auto_corr,eq:fw_cross_corr},
            it is easy to see that
            \begin{align}\label{eq:auto_equals_cross_finwidth}
                \begin{split}
                    \iint\!\dd q\,\dd q'\auto\of{q,q'} &= \iint\!\dd q\,\dd q'\cross\of{q,q'} \\
                        &=\sum_{n} \Lambda_{n} + \sum_{n} \Lambda_{n}^2.
                \end{split}
            \end{align}
            Therefore,  if the transfer functions remain unchanged,
            the covariance is only increased by a factor of $2$ for completely indistinguishable  signal and idler
            photons.

        \subsection{Plane-wave pump}\label{sec:integ_cov_deg_case_plane_wave}
            In the case of a plane-wave pump, it is possible to derive similar relations to
            \Cref{eq:eta_beta_rels_from_comm_bbee,eq:eta_beta_rels_from_comm_ebeb}
            by using the fact that the solutions~\eqref{eq:sol_ops_pw_both} must also fulfill the bosonic commutation relations.
            This leads to the following relations  between $\tilde{\eta}_{\mathrm{pw}}$ and
            $\beta_{\mathrm{pw}}$:
            \begin{subequations}
                \begin{align}
                    \left|\tilde{\eta}_{\mathrm{pw}}\of{q_s}\right|^2 &=
                        1 + \left|\beta_{\mathrm{pw}}\of{q_s}\right|^2, \\
                    \tilde{\eta}_{\mathrm{pw}}\of{q_s} \beta_{\mathrm{pw}}\of{-q_s} &=
                        \tilde{\eta}_{\mathrm{pw}}\of{-q_s} \beta_{\mathrm{pw}}\of{q_s}.
                \end{align}
            \end{subequations}
            These relations have also been found in
            Refs.~\cite{PhysRevA.69.023802,Brambilla2001,RevModPhys.71.1539,klyshko1988photons,PhysRevA.98.053827}
            and have been used to obtain
            \Cref{eq:cov_pw_cc,eq:pw_cov_general_finite_spectral} and
            \Cref{eq:pw_auto_corr,eq:pw_cross_corr,eq:cov_pw_finite_deltaqsqr} below.
            \par As in the finite-width case above, the covariance for the fully
            degenerate case can again be split as
            \begin{subequations}
                \begin{align}
                    \cov_{\mathrm{pw}}\of{q,q'} = \auto_{\mathrm{pw}}\of{q,q'} + \cross_{\mathrm{pw}}\of{q,q'},
                \end{align}%
                where%
                \begin{align}
                    \begin{split}
                        \auto_{\mathrm{pw}}\of{q,q'} &= \delta\of{q-q'} \NN_s\of{q}
                        \left[ 1 + \NN_s\of{q} \right], \label{eq:pw_auto_corr}
                    \end{split} \\
                    \cross_{\mathrm{pw}}\of{q,q'} &= \delta\of{q+q'}
                        \NN_s\of{q} \left[ 1 +
                         \left|\beta_{\mathrm{pw}}\of{-q}\right|^2 \right]. \label{eq:pw_cross_corr}
                \end{align}
            \end{subequations}
            \par Alternatively to
            \Cref{eq:diffeqs_eta_beta_2_pw_s,eq:diffeqs_eta_beta_2_pw_i}, the system of
            differential equations can also be obtained in the form
            \begin{subequations}
                \begin{align}
                    \frac{\dd \beta_{\mathrm{pw}}\of{-q_s,L}}{\dd L} &= 
                        \Gamma_0 h\of{q_s,L} \tilde{\eta}^{*}_{\mathrm{pw}}\of{q_s,L},\\
                    \frac{\dd \tilde{\eta}^{*}_{\mathrm{pw}}\of{q_s,L}}{\dd L} &= 
                        \Gamma_0 h^{*}\of{q_s,L} \beta_{\mathrm{pw}}\of{-q_s,L},
                \end{align}
            \end{subequations}
            which, evidently, are equivalent to \Cref{eq:diffeqs_eta_beta_2_pw_s,eq:diffeqs_eta_beta_2_pw_i}
            except that $q_s\leftrightarrow-q_s$ is swapped in the  transfer functions
            $\tilde{\eta}^{*}_{\mathrm{pw}}$ and $\beta_{\mathrm{pw}}$.
            Note that
            this system can also be obtained from
            \Cref{eq:diffeqs_eta_beta_2_pw_s,eq:diffeqs_eta_beta_2_pw_i}
            directly by using the fact that $h$ is an even function of $q_s$.
            \par Ultimately, this implies that the transfer functions are also even functions with respect to the
            wave vector:
            \begin{subequations}
                \begin{align}
                    \beta_{\mathrm{pw}}\of{q,L} &= \beta_{\mathrm{pw}}\of{-q,L}, \label{eq:beta_even_in_q} \\
                    \tilde{\eta}^{*}_{\mathrm{pw}}\of{q,L} &= \tilde{\eta}^{*}_{\mathrm{pw}}\of{-q,L},
                \end{align}
            \end{subequations}
            and therefore, the intensity spectrum is also the even function with respect to the wave vector.
            \par Combining \Cref{eq:pw_cross_corr} and \Cref{eq:beta_even_in_q}
            leads to the following expression fro the cross-correlation term:
            \begin{align}
                \cross_{\mathrm{pw}}\of{q,q'} = \delta\of{q+q'} \NN_s\of{q} \left[ 1 + \NN_s\of{q} \right].
            \end{align}
            According to \Cref{eq:pw_auto_corr} this means that
            \begin{align}
                \auto_{\mathrm{pw}}\of{q,q'} &= \cross_{\mathrm{pw}}\of{q,-q'},
            \end{align}
            and, finally,
            \begin{align}
                \iint\!\dd q\,\dd q'\auto_{\mathrm{pw}}\of{q,q'} &= \iint\!\dd q\,\dd q'\cross_{\mathrm{pw}}\of{q,q'}.
            \end{align}
            Therefore, similarly to the finite-width pump case [\Cref{eq:auto_equals_cross_finwidth}],
            the integral covariance
            is only scaled up by a factor of $2$ when the signal and idler
            photons are indistinguishable.

    \section{Solution of the integro-differential equations for different interferometer phases}
            \label{sec:integ_diffeq_sol_details}
        Throughout this work, it is often required to obtain 
        solutions for the  system of  integro-differential
         equations ~\eqref{eq:integ_diffeqs_both}
        or~\eqref{eq:integ_diffeqs_eta_beta_both} for many
        different interferometer phases $\phi$. To simplify this process and drastically reduce
        the numerical complexity, it is possible to split the solution process into two steps.
        \par First, the system~\eqref{eq:integ_diffeqs_eta_beta_both}
        is solved for the first crystal of the interferometer,
        yielding functions $\eta^{\left(1\right)}$ and $\beta^{\left(1\right)}$ connecting the output
        plane-wave operators of the first crystal $\hat{a}_{s/i}^{\left(1,\mathrm{out}\right)}$ to the vacuum plane-wave operators:
        \begin{subequations}
            \begin{align}
                \begin{split}
                    \hat{a}_s^{\left(1,\mathrm{out}\right)}\of{q_s} =\ & \int\!\dd
                        q_s'\,\tilde{\eta}^{\left(1\right)}\of{q_s,q_s'} \hat{a}_s\of{q_s'} \\
                    &+ \int\!\dd q_i'\,\beta^{\left(1\right)}\of{q_s,q_i'} \hat{a}_i^{\dagger}\of{q_i'},
                    \label{eq:sol_ops_first_crystal_s}
                \end{split} \\
                \begin{split}
                    \left[\hat{a}_i^{\left(1,\mathrm{out}\right)}\of{q_i}\right]^{\dagger} =\ & \int\!\dd q_i'\,
                        \left[\tilde{\eta}^{\left(1\right)}\of{q_i,q_i'}\right]^{*} \hat{a}_i^{\dagger}\of{q_i'} \\
                        &+ \int\!\dd q_s'\,\left[\beta^{\left(1\right)}\of{q_i,q_s'}\right]^{*} \hat{a}_s\of{q_s'},
                    \label{eq:sol_ops_first_crystal_i}
                \end{split}
            \end{align}
        \end{subequations}
        where for simplicity, we have dropped the length~$L$ and the frequency dependence~$\omega$
        from the arguments of the operators.
        The index~$^{\left(1\right)}$ indicates that these functions are the solutions for
        the first crystal.
        The initial conditions for the functions read
        \begin{subequations}
            \begin{align}
                \beta^{\left(1\right)}\of{q,q',L=0} &= 0, \\
                \left[\tilde{\eta}^{\left(1\right)}\of{q,q',L=0}\right]^{*} &= \delta\of{q-q'}.
            \end{align}
        \end{subequations}
        \par Next, we can set up a similar set of equations for the second crystal:
        \begin{subequations}
            \begin{align}
                \begin{split}
                    \hat{a}_s^{\left(2,\mathrm{out}\right)}\of{q_s} =\ & \int\!\dd q_s'\,
                        \tilde{\eta}_{\phi}^{\left(2\right)}\of{q_s,q_s'} \hat{a}_s^{\left(2,\mathrm{in}\right)}\of{q_s'} \\
                    &+ \int\!\dd q_i'\,\beta_{\phi}^{\left(2\right)}\of{q_s,q_i'}
                        \left[\hat{a}_i^{\left(2,\mathrm{in}\right)}\of{q_i'}\right]^{\dagger},
                    \label{eq:sol_ops_second_crystal_no_phase_s}
                \end{split}\\
                \begin{split}
                    \left[\hat{a}_i^{\left(2,\mathrm{out}\right)}\of{q_i}\right]^{\dagger} =\ &
                        \int\!\dd q_i'\,
                        \left[\tilde{\eta}_{\phi}^{\left(2\right)}\of{q_i,q_i'}\right]^{*}
                        \left[\hat{a}_i^{\left(2,\mathrm{in}\right)}\of{q_i'}\right]^{\dagger} \\
                    &+ \int\!\dd q_s'\,\left[\beta_{\phi}^{\left(2\right)}\of{q_i,q_s'}\right]^{*}
                        \hat{a}_s^{\left(2,\mathrm{in}\right)}\of{q_s'},
                    \label{eq:sol_ops_second_crystal_no_phase_i}
                \end{split}
            \end{align}
        \end{subequations}
        which now connects the input operators of the second crystal (not the vacuum operators)
        to its output operators. We have added the index $_{\phi}$ to emphasize that the solution
        of this system depends on the interferometer phase $\phi$ via the function $h^{\left(2\right)}$
        describing the second crystal, see \Cref{eq:h_noncomp,eq:h_comp}.
        \par Plugging this back into the
        system of integro-differential
        equations~\eqref{eq:integ_diffeqs_both}
        yields, again, the same system of
        integro-differential equations~\eqref{eq:integ_diffeqs_eta_beta_both} with $\tilde{\eta}$ and $\beta$ replaced by
        $\tilde{\eta}^{\left(2\right)}_{\phi}$ and $\beta^{\left(2\right)}_{\phi}$,
        respectively. This
        system of the integro-differential equations
        is solved for $L\in\left[0,L_1\right]$,
        where $L_1$ is the crystal length,
        and
        has the same initial conditions as for
        the first crystal:
        \begin{subequations}
            \begin{align}\label[pluralequation]{eq:initial_values_second_crystal}
                \beta_{\phi}^{\left(2\right)}\of{q,q',L=0} &= 0, \\
                \left[\tilde{\eta}_{\phi}^{\left(2\right)}\of{q,q',L=0}\right]^{*} &= \delta\of{q-q'}.
            \end{align}
        \end{subequations}
        \par The phase matching function $h^{\left(2\right)}$ which describes the second crystal can be
        factorized as
        \begin{align}
            h^{\left(2\right)}\of{q_s,q_i,L} &= p^{\left(2\right)}\of{q_s,q_i,L} \E^{\I \phi},
        \end{align}
        where $p^{\left(2\right)}$ is a complex-valued function.
        \par Plugging this form of $h^{\left(2\right)}$ into the system of integro-differential equations~\eqref{eq:integ_diffeqs_eta_beta_both} and 
        defining
        \begin{align}
            \beta^{\left(2\right)}\of{q,q'} &\eqdef \beta_{\phi}^{\left(2\right)}\of{q,q'} \E^{-\I \phi},\\
            \tilde{\eta}^{\left(2\right)}\of{q,q'}&\eqdef \tilde{\eta}_{\phi}^{\left(2\right)}\of{q,q'},
        \end{align}
        one can obtain the following system of integro-differential equations for the second crystal:
        \begin{widetext}
            \begin{subequations}
                \begin{align}
                    \frac{\dd \beta^{(2)}\of{q_s,q_i',L}}{\dd L} &=\Gamma\int\!\dd q_i\,
                        \E^{-\frac{\left(q_s+q_i\right)^2\sigma^2}{2}}
                    p^{\left(2\right)}\of{q_s,q_i,L} [\tilde{\eta}^{(2)}\of{q_i,q_i',L}]^{*}, \\
                    \frac{\dd [\tilde{\eta}^{(2)}\of{q_i,q_i',L}]^{*}}{\dd L} &=\Gamma\int\!\dd q_s\,
                        \E^{-\frac{\left(q_s+q_i\right)^2\sigma^2}{2}}
                    \left[p^{\left(2\right)}\of{q_s,q_i,L}\right]^{*} \beta^{(2)}\of{q_s,q_i',L},
                \end{align}
            \end{subequations}
           where  the initial conditions for the transfer functions are given by
            \begin{subequations}
                \begin{align}
                    \beta^{\left(2\right)}\of{q,q',L=0} &= 0, \\
                    \left[\tilde{\eta}^{\left(2\right)}\of{q,q',L=0}\right]^{*} &= \delta\of{q-q'}.
                \end{align}
            \end{subequations}
        \end{widetext}
        This system, importantly, results in a solution for 
        $\tilde{\eta}^{\left(2\right)}$ and $\beta^{\left(2\right)}$ that is independent of $\phi$.
        \par The solution of the integro-differential equations for the second crystal now reads
        \begin{widetext}
            \begin{subequations}
                \begin{align}
                    \hat{a}_s^{\left(2,\mathrm{out}\right)}\of{q_s} &= \int\!\dd q_s'\,
                        \tilde{\eta}^{\left(2\right)}\of{q_s,q_s'} \hat{a}_s^{\left(2,\mathrm{in}\right)}\of{q_s'}
                        + \E^{\I \phi} \int\!\dd q_i'\,\beta^{\left(2\right)}\of{q_s,q_i'}
                        \left[\hat{a}_i^{\left(2,\mathrm{in}\right)}\of{q_i'}\right]^{\dagger}, \\
                    \left[\hat{a}_i^{\left(2,\mathrm{out}\right)}\of{q_i}\right]^{\dagger} &= \int\!\dd q_i'\, \left[\tilde{\eta}^{\left(2\right)}\of{q_i,q_i'}\right]^{*}
                        \left[\hat{a}_i^{\left(2,\mathrm{in}\right)}\of{q_i'}\right]^{\dagger}
                        + \E^{-\I \phi} \int\!\dd q_s'\,\left[\beta^{\left(2\right)}\of{q_i,q_s'}\right]^{*}
                        \hat{a}_s^{\left(2,\mathrm{in}\right)}\of{q_s'}.
                \end{align}
            \end{subequations}
        \end{widetext}
        \par The full solution for the entire \suoo interferometer 
        consisting of two crystals is defined via
        \begin{subequations}
            \begin{align}
                \begin{split}
                    \hat{a}_s^{\left(\mathrm{SU},\mathrm{out}\right)}\of{q_s} =\ & \int\!\dd
                        q_s'\,\tilde{\eta}^{\left(\mathrm{SU}\right)}\of{q_s,q_s'}
                        \hat{a}_s\of{q_s'} \\
                    &+ \int\!\dd q_i'\,\beta^{\left(\mathrm{SU}\right)}\of{q_s,q_i'}
                    \hat{a}_i^{\dagger}\of{q_i'},
                \end{split}\\
                \begin{split}
                    \left[\hat{a}_i^{\left(\mathrm{SU},\mathrm{out}\right)}\of{q_i}\right]^{\dagger}
                    =\ & \int\!\dd q_i'\,
                        \left[\tilde{\eta}^{\left(\mathrm{SU}\right)}\of{q_i,q_i'}\right]^{*}
                        \hat{a}_i^{\dagger}\of{q_i'} \\
                    &+ \int\!\dd
                        q_s'\,\left[\beta^{\left(\mathrm{SU}\right)}\of{q_i,q_s'}\right]^{*}
                        \hat{a}_s\of{q_s'},
                \end{split}
            \end{align}
        \end{subequations}
        where the functions $\tilde{\eta}^{\left(\mathrm{SU}\right)}$ and $\beta^{\left(\mathrm{SU}\right)}$
        connect the output operators of the entire interferometer $\hat{a}_{s/i}^{\left(\mathrm{SU},\mathrm{out}\right)}$ with the vacuum plane-wave operators.
        Furthermore, note that
        \begin{subequations}
            \begin{align}
                \hat{a}_s^{\left(2,\mathrm{in}\right)} &= \hat{a}_s^{\left(1,\mathrm{out}\right)}, \\
                \hat{a}_i^{\left(2,\mathrm{in}\right)} &= \hat{a}_i^{\left(1,\mathrm{out}\right)}.
            \end{align}
        \end{subequations}
        Using this equality between the output plane-wave operators 
        of the first crystal and the input plane-wave operators 
        of the second crystal to connect
        \Cref{eq:sol_ops_first_crystal_s,eq:sol_ops_first_crystal_i}
        with \Cref{eq:sol_ops_second_crystal_no_phase_s,eq:sol_ops_second_crystal_no_phase_i}, one can obtain the following 
       connection relations between the transfer functions of the entire interferometer and each crystal:
        \begin{subequations}
            \begin{align}
                \begin{split}
                    \tilde{\eta}^{\left(\mathrm{SU}\right)}\of{q,q'} &= \int\!\dd\bar{q}\,
                        \tilde{\eta}_{\phi}^{\left(2\right)}\of{q,\bar{q}} \tilde{\eta}^{\left(1\right)}\of{\bar{q},q'} \\
                    &\qquad+ \int\!\dd\bar{q}\,\beta_{\phi}^{\left(2\right)}\of{q,\bar{q}}
                        \left[\beta^{\left(1\right)}\of{\bar{q},q'}\right]^{*},
                    \label{eq:composite_tf_eta_nophase}
                \end{split}\\
                \begin{split}
                    \beta^{\left(\mathrm{SU}\right)}\of{q,q'} &= \int\!\dd\bar{q}\,
                        \tilde{\eta}_{\phi}^{\left(2\right)}\of{q,\bar{q}} \beta^{\left(1\right)}\of{\bar{q},q'} \\
                    &\qquad+ \int\!\dd\bar{q}\,\beta_{\phi}^{\left(2\right)}\of{q,\bar{q}}
                        \left[\tilde{\eta}^{\left(1\right)}\of{\bar{q},q'}\right]^{*}.
                        \label{eq:composite_tf_beta_nophase}
                \end{split}
            \end{align}
        \end{subequations}
        Alternatively, including the interferometer phase, the relations above read:
        \begin{subequations}
            \begin{align}
                \begin{split}
                    \tilde{\eta}^{\left(\mathrm{SU}\right)}\of{q,q'} &= \int\!\dd\bar{q}\,
                        \tilde{\eta}^{\left(2\right)}\of{q,\bar{q}} \tilde{\eta}^{\left(1\right)}\of{\bar{q},q'} \\
                    &\qquad+ \E^{\I\phi}\int\!\dd\bar{q}\,\beta^{\left(2\right)}\of{q,\bar{q}}
                        \left[\beta^{\left(1\right)}\of{\bar{q},q'}\right]^{*},
                    \label{eq:composite_tf_eta_phase}
                \end{split}\\
                \begin{split}
                    \beta^{\left(\mathrm{SU}\right)}\of{q,q'} &= \int\!\dd\bar{q}\,
                        \tilde{\eta}^{\left(2\right)}\of{q,\bar{q}} \beta^{\left(1\right)}\of{\bar{q},q'} \\
                    &\qquad+ \E^{\I\phi}\int\!\dd\bar{q}\,\beta^{\left(2\right)}\of{q,\bar{q}}
                        \left[\tilde{\eta}^{\left(1\right)}\of{\bar{q},q'}\right]^{*}.
                    \label{eq:composite_tf_beta_phase}
                \end{split}
            \end{align}
        \end{subequations}
        \par Hence, it is possible to solve the set of
        the integro-differential equations independently for
        both crystals. Using $p^{\left(2\right)}$ instead of $h^{\left(2\right)}$ for
        the second crystal makes it possible to obtain
        a phase independent solution for the second crystal.
        Then, the phase independent solutions of both crystals
        can be combined for an arbitrary value of $\phi$ according to
        \Cref{eq:composite_tf_eta_phase,eq:composite_tf_beta_phase} without
        having to solve the integro-differential equations
        for each value of phase.
    
    \section{Behavior of the fitting constant \texorpdfstring{$A$}{A} with increasing gain}\label{sec:A_G_dependence}
        As it was shown in \Cref{sec:pw_intro}, for the plane-wave pump, 
        the collinear intensity behaves  as~$\propto\sinh^2\of{G}$,
        where~$G=\Gamma L_1$ is the parametric gain and $L_1$ is
        the crystal length. Similarly,  the $\sinh^2$-behavior
        is expected for the collinear emission in the case of pumping with an arbitrary width~\cite{PRR2}.
        However, in general, for the multimode light, this tendency is not preserved for all gain intervals, namely, the fitting parameter is no longer a constant over the full gain range.
        \par In order to obtain the correct gain-dependence, we therefore perform
        a series of fits of collinear intensities, one for each small interval of theoretical gain~$\Gamma$ under
        consideration.
        From this procedure we effectively obtain the fitting parameter
        $A_{\Gamma}$ for different values of $\Gamma$. Then, the experimental
        gain is given by $G=A_{\Gamma}\Gamma$. We have added
        a subscript~$_{\Gamma}$ to show that  the
        fit parameter~$A$ now depends on~$\Gamma$.
        Thus, the fitting formula now reads
        $y\of{\Gamma}\propto\sinh^2\of{A_{\Gamma}\Gamma}$.
        \par
\begin{figure}[tb]
            \includegraphics[width=\linewidth]{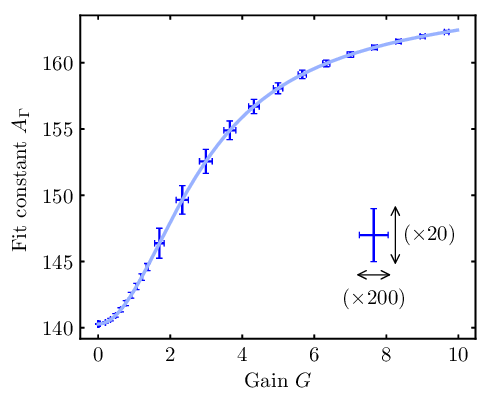}%
            \caption{Values of the fitting parameter $A$ versus the parametric
                gain. The error bars correspond to the standard error of
                the fitting procedure. The bars are scaled up by a factor
                of~$200$ along~$G$ and by a factor of~$20$ along~$A$ for better
                visibility. Note that the density of points is higher for
                ${G<1.9}$ that results in decrease of the error below
                ${G=1.9}$. See also \Cref{tab:G_and_A_of_Gamma} for the selected
                values.\label{fig:A_of_G}}%
        \end{figure}

 The values of~$A_{\Gamma}$ corresponding to the experimental
        gains  used for the plots shown in \Cref{sec:finwidth}
        are given in \Cref{tab:G_and_A_of_Gamma}. Furthermore,
        we show the full behavior of $A$ over an entire range of
        gain values in \Cref{fig:A_of_G}.
        It can be seen that in the limit of high gain values (when the radiation tends to a single mode), the fitting parameter approaches the constant value.
        The uncertainties indicated for $A_{\Gamma}$
        and~$G$ are the uncertainties introduced by the fitting
        procedure. Note that since the fit parameter $A_{\Gamma}$
        has an uncertainty assigned,~$G$ can also  be provided 
        only within some uncertainty.

        \begin{table}
            \renewcommand{\arraystretch}{1.3}
        	\begin{center}
        		\caption{Fitting parameter $A$ for the selected gains $G$.\label{tab:G_and_A_of_Gamma}}
        		\begin{tabularx}{\linewidth}{YY}
                    \hline\hline
        			Gain $G$ & Fit constant $A_{\Gamma}$ \\
        			\hline
        			$0.01\pm 5\times 10^{-11}$ & $140.285\pm 6\times 10^{-7}$ \\
        			$1.25\pm 1.2\times 10^{-4}$ & $144.029\pm 0.013$ \\
        			$2.5\pm 9\times 10^{-4}$ & $150.44\pm 0.06$ \\
        			$3.75\pm 9\times 10^{-4}$ & $155.19\pm 0.04$ \\
        			$5.0\pm 7\times 10^{-4}$ & $158.080\pm 0.021$ \\
        			\hline\hline
        		\end{tabularx}
        	\end{center}
        \end{table}
        
    \section{Divergence in the plane-wave pump case}\label{sec:divergence_treatment}
        \par After quantizing the electromagnetic field in a cavity of 
        length $\El_x$ in terms of the discrete plane-wave
        operators~$\hat{a}_{q}$ obeying the commutation
        relations~\cite{gerry_knight_2004}
        \begin{align}
            [\hat{a}_{q}, \hat{a}_{q'}^{\dagger}] &= \delta_{q,q'},
        \end{align}
        it is common to take the
        \textit{continuous limit} ${\El_x\to\infty}$, which
        allows the transition from the discretely spaced
        wave vectors with spacing $\Delta q=\frac{2\pi}{\El_x}$ 
        to a continuous spectrum
        (${\Delta q\to 0}$)~\cite{PhysRevA.42.4102}.
        \par Here, we have assumed a fixed frequency for the cavity
        modes of interest. This is why we only consider
        the commutator in the transverse variables, since the
        $z$-components of the wave vectors are strictly fixed by given $q$ and the frequency.
        \par Formally, the transition to the continuous limit is performed by including the
        factor $\sqrt{\Delta q}$ in the discrete plane-wave operators $\hat{a}_{q}$
        and introducing new operators ${\hat{a}\of{q}=\hat{a}_{q} / \sqrt{\Delta q}}$.
        \par For the PDC process, the commutation relations of the newly defined plane-wave operators
        are then given by
        \begin{subequations}
            \begin{align}
                [\hat{a}_s\of{q_s,L,\omega_s}, \hat{a}_s^{\dagger}\of{q_s',L,\omega_s}] &= \frac{\delta_{q_s,q_s'}}{\Delta q},
                \label{eq:comms_discrete_ss}\\
                [\hat{a}_i\of{q_i,L,\omega_i}, \hat{a}_i^{\dagger}\of{q_i',L,\omega_i}] &= \frac{\delta_{q_i,q_i'}}{\Delta q},
                \label{eq:comms_discrete_ii}\\
                [\hat{a}_s\of{q_s,L,\omega_s}, \hat{a}_i^{\dagger}\of{q_i',L,\omega_i}] &= 0.
                \label{eq:comms_discrete_si}
            \end{align}
        \end{subequations}
        Note that these commutation relations are the discrete analogues of the ones
        given in \Crefrange{eq:commrel_ss_finwidth}{eq:commrel_si_finwidth}. In fact, when taking the continuous limit,
        the commutators from
        \Crefrange{eq:comms_discrete_ss}{eq:comms_discrete_si} approach
        \Crefrange{eq:commrel_ss_finwidth}{eq:commrel_si_finwidth},
        respectively~\cite{gerry_knight_2004,PhysRevA.42.4102}.
        \par As was seen in \Cref{sec:pw_intro}, taking the continuous limit
        for the plane-wave pump
        approximation
        results in a divergence
        of the intensity spectrum, namely,
        $\langle\hat{N}_s\of{q_s}\rangle = {{\langle 0|\,\hat{a}_s^{\dagger}
        \of{q_s}\hat{a}_s\of{q_s}|0\rangle}\propto\delta\of{0}}$.
        Physically, if the quantization size of the system 
        becomes unbounded while the system is pumped by a plane wave covering all of space,
        an infinite amount of PDC photons is created along the transverse dimension.
        \par To avoid this divergence and make the description more mathematically rigorous, it is possible
        to repeat the derivation for the integro-differential equations presented in Ref.~\cite{PRR2}
        with a plane-wave pump for the discrete operators.
        In that case, the integrals over wave vectors
        have to be replaced with the sums over wave vectors
        with the correspondence
        ${\dd q \leftrightarrow \Delta q}$.
        The system which generates the PDC radiation then 
        consists of the quantization box with the periodic
        boundary conditions and is again described by the
        system of differential
        equations~\eqref{eq:diffeqs_eta_beta_2_pw_both} with 
        the form of the solution~\eqref{eq:sol_ops_pw_both}.
        \par In this case, evaluating the expectation
        value in \Cref{eq:intens_pw_delta} yields
        \begin{align}
            \langle\hat{N}_s\of{q_s}\rangle &= \frac{1}{\Delta q}
                \left|\beta_{\mathrm{pw}}\of{q_s,L}\right|^2.
        \end{align}
        Clearly, the quantity defined by
        \begin{align}\label{eq:intens_dens_pw}
            \NN_s\of{q_s} &\eqdef \Delta q\,\langle\hat{N}_s\of{q_s}\rangle = \left|\beta_{\mathrm{pw}}\of{q_s,L}\right|^2
        \end{align}
        describes the intensity per length $\frac{\El_x}{2\pi}$ of the system.
        The important observation is now that $\NN_s\of{q_s}$ remains
        unchanged and
        finite when taking the continuous limit
        $\El_x\to\infty$ (that is, it is independent of~$\El_x$).
        \par Similarly, the covariance evaluated from its definition [\Cref{eq:def_cov}]
        is given by
        \begin{align}\label{eq:cov_pw_finite_deltaqsqr}
            \cov\of{q_s,q_s'} &= \frac{\delta_{q_s,q_s'}}{\left(\Delta q\right)^2}\,\NN_s\of{q_s} \left[ 1 +
                \NN_s\of{q_s}\right].
        \end{align}
        After defining
        \begin{align}
            \cov_\mathrm{pw}\of{q_s,q_s'} &\eqdef \Delta q \cov\of{q_s,q_s'},
        \end{align}
        the continuous limit can be taken ($\El_x\to\infty$)
        and results in the well-behaved quantity
        \begin{align}\label{eq:cov_dens_pw}
            \cov_\mathrm{pw}\of{q_s,q_s'} &= \NN_s\of{q_s} \left[ 1 +
                \NN_s\of{q_s}\right] \delta\of{q_s-q_s'},
        \end{align}
        which corresponds to the covariance per the transverse length
        of the system.
        \par In order to make the results for the plane-wave and the finite-width
        pump comparable, we consider such a transverse size $L_x$ of the
        system in the case of plane-wave pumping, which gives the same
        integral intensity after the first crystal as in the
        finite-width pump case. Note that $L_x$ is different to the
        quantization length $\El_x$. Then, 
        \begin{align}\label{eq:length}
            L_x = \frac{\langle\hat{N}_{s,\mathrm{tot}}\rangle}{\NN_{s,\mathrm{tot}}},
        \end{align}
        where $\langle\hat{N}_{s,\mathrm{tot}}\rangle$ is the integral intensity
        in the case of the finite-width pump and 
        \begin{align}\label{eq:integral_phot_dens_pw}
            \NN_{s,\mathrm{tot}} &= \int\!\dd q_s\,\NN_{s}\of{q_s}
        \end{align}
        is the integral photon density (integral intensity per transverse length),
        analogously to \Cref{eq:def_Ntot_op}.
        \par Note that this step is not necessary for the SNL-normalized phase sensitivity $f$, since according to
        \Cref{eq:def_f,eq:def_delta_phi}, it does not depend on the factor $L_x$. Indeed, obtaining the integral number of photons and covariance by
        multiplying their densities [\Cref{eq:cov_dens_pw,eq:intens_dens_pw}] 
        by $L_x$ and substituting them into \Cref{eq:def_f,eq:def_delta_phi}
        shows that the additional factors $L_x$ are cancelled out.

    \section{Analytical implications of the compensation}\label{sec:analytic_impl_comp}
        More explicitly, the solutions of the integro-differential equations
        [\Cref{eq:as_dagger_sol,eq:ai_dagger_sol}] can be written as
        \begin{subequations}
            \label{eq:sol_ops_LL_both}
            \begin{align}
                \begin{split}
                    \hat{a}_s\of{q_s,L_1} =& \int\!\dd q_s'\,\tilde{\eta}\of{q_s,q_s';L_1,L_0} \hat{a}_s\of{q_s',L_0} \\
                    &+ \int\!\dd q_i'\,\beta\of{q_s,q_i';L_1,L_0} \hat{a}_i^{\dagger}\of{q_i',L_0},
                    \label{eq:sol_ops_LL_s}
                \end{split}\\
                \begin{split}
                    \hat{a}_i^{\dagger}\of{q_i,L_1} =& \int\!\dd q_i'\,
                        \tilde{\eta}^{*}\of{q_i,q_i';L_1,L_0} \hat{a}_i^{\dagger}\of{q_i',L_0} \\
                        &+ \int\!\dd q_s'\,\beta^{*}\of{q_i,q_s';L_1,L_0} \hat{a}_s\of{q_s',L_0},
                        \label{eq:sol_ops_LL_i}
                \end{split}
            \end{align}
        \end{subequations}
        which corresponds to a more general case where the transfer functions
        $\tilde{\eta}\of{q,q';L_1,L_0}$ and $\beta\of{q,q';L_1,L_0}$ connect
        the plane-wave operators at~$L_0$ to the ones at~$L_1$.
        \par The inverse transform of the solutions~\eqref{eq:sol_ops_LL_both} is given
        by~\cite{Christ_2013,PhysRevA.43.3934,PhysRevA.71.055801}:
        \begin{subequations}
            \label{eq:sol_ops_LL_inverse_both}
            \begin{align}
                \begin{split}
                    \hat{a}_s\of{q_s,L_0} =& \int\!\dd q_s'\,\tilde{\eta}^{*}\of{q_s',q_s;L_1,L_0} \hat{a}_s\of{q_s',L_1} \\
                    &- \int\!\dd q_i'\,\beta\of{q_i',q_s;L_1,L_0} \hat{a}_i^{\dagger}\of{q_i',L_1},
                    \label{eq:sol_ops_LL_inverse_s}
                \end{split}\\
                \begin{split}
                    \hat{a}_i^{\dagger}\of{q_i,L_0} =& \int\!\dd q_i'\,
                        \tilde{\eta}\of{q_i',q_i;L_1,L_0} \hat{a}_i^{\dagger}\of{q_i',L_1} \\
                        &- \int\!\dd q_s'\,\beta^{*}\of{q_s',q_i;L_1,L_0} \hat{a}_s\of{q_s',L_1},
                        \label{eq:sol_ops_LL_inverse_i}
                \end{split}
            \end{align}
        \end{subequations}
        which can be immediately verified by plugging
        the form of the solutions~\eqref{eq:sol_ops_LL_inverse_both}
        into the solutions~\eqref{eq:sol_ops_LL_both} and
        using \Cref{eq:eta_beta_rels_from_comm_bbee,eq:eta_beta_rels_from_comm_ebeb}.
        \par For completeness, we would like to mention that by
        applying the constraint
        that $\hat{a}_s\of{q_s,L_0}$ and $\hat{a}_i^{\dagger}\of{q_i,L_0}$
        in the solutions~\eqref{eq:sol_ops_LL_inverse_both} have to
        obey the bosonic commutation relations,
        one can obtain two additional relationships similar to
        \Cref{eq:eta_beta_rels_from_comm_bbee,eq:eta_beta_rels_from_comm_ebeb}.
        Using
        \Cref{eq:commrel_ss_finwidth,eq:commrel_ii_finwidth},
        one can obtain
        \begin{subequations}
            \begin{align}
                \begin{split}
                    &\int\!\dd \bar{q}\, \beta\of{\bar{q},q,L}\beta^{*}\of{\bar{q},q',L} + \delta\of{q-q'}\\
                        &\qquad=\int\!\dd \bar{q}\, \tilde{\eta}^{*}\of{\bar{q},q,L}\tilde{\eta}\of{\bar{q},q',L},
                \end{split}
            \end{align}
           and using \Cref{eq:commrel_si_finwidth_nodagger}, one obtains
            \begin{align}
                \int\!\dd \bar{q}\, \tilde{\eta}^{*}\of{\bar{q},q,L}\beta\of{\bar{q},q',L} 
                   = \int\!\dd \bar{q}\, \beta\of{\bar{q},q,L}\tilde{\eta}^{*}\of{\bar{q},q',L}.
            \end{align}
        \end{subequations}
        These have also been found in Ref.~\cite{Christ_2013} for a more general PDC process
        in the frequency domain.
        \par For a single crystal ranging
        from the coordinates~$L_0=0$ (beginning of the crystal)
        and~$L_1$ (end of the crystal), the system of the
        integro-differential equations for the
        inverse transform [\Cref{eq:sol_ops_LL_inverse_s,eq:sol_ops_LL_inverse_i}]
        is given by:
        \begin{widetext}
            \begin{equation}
                \begin{gathered}\label{eq:system_single_crystal}
                    \int_{L_1}^{L_0}\!\dd L\begin{cases}
                        \displaystyle \frac{\dd \beta^{\left(1\right)}\of{q_i',q_s;L_1,L}}{\dd L} = -\Gamma \int\!\dd q_i\,
                            r\of{q_s,q_i,L} \E^{\I \Delta k L} \tilde{\eta}^{\left(1\right)}\of{q_i',q_i;L_1,L} \\
                        \displaystyle \frac{\dd \tilde{\eta}^{\left(1\right)}\of{q_i',q_i;L_1,L}}{\dd L} = 
                            -\Gamma \int\!\dd q_s\,
                        r^{*}\of{q_s,q_i,L} \E^{-\I \Delta k L}\beta^{\left(1\right)}\of{q_i',q_s;L_1,L} \end{cases}\\
                   \begin{aligned}
                        \tilde{\eta}^{\left(1\right)}\of{q_i',q_i;L_1,L_1} &= \delta\of{q_i-q_i'} \\
                        \beta^{\left(1\right)}\of{q_i',q_s;L_1,L_1} &= 0.
                   \end{aligned}
                \end{gathered}
            \end{equation}
        \end{widetext}
        Here we have assumed a more general case than in the main text, where $r\of{q_s,q_i,L}$
        contains the pump term and might additionally depend on the integration coordinate $L$.
        Note that the integration interval for the solution of the system starts at
        the upper bound $L_1$ of 
        the PDC section. The initial
        value conditions state that the plane-wave operators are known at $L_1$. After
        performing the integration, the obtained functions
        $\beta^{\left(1\right)}\of{q_i',q_s;L_1,L_0}$ and
        $\tilde{\eta}^{\left(1\right)}\of{q_i',q_i;L_1,L_0}$ connect
        the plane-wave operators at~$L_1$ to
        the ones at the beginning of the crystal at~$L_0$.
        \par Next, we consider the second crystal for the 
        compensated configuration with the same length
        as the first crystal ranging from~$L_1$
        to ${L_2=2L_1}$. As  shown in
        \Cref{sec:integ_diffeq_sol_details}, it 
        is not necessary to solve the system
        of integro-differential equations with the phase term $\E^{\I\phi}$ included in
        $r^{\left(2\right)}\of{q_s,q_i,L}$. Instead, one can first integrate
        the equations without the phase and then multiply $\beta^{\left(2\right)}$ with the phase
        afterwards to obtain the solution for the second crystal including the phase.
        Furthermore, note that we will assume that ${r^{\left(2\right)}\of{q_s,q_i,L}=r\of{q_s,q_i,L}}$.
        As will be seen below, this is a necessary requirement to obtain compensation.
        The system of the integro-differential equations and the initial
        value conditions for the phase independent transfer functions of the second crystal read:
        \begin{widetext}
            \begin{equation}
                \begin{gathered}
                    \int_{L_1}^{L_2}\!\dd L\begin{cases}
                        \displaystyle \frac{\dd \beta^{\left(2\right)}\of{q_s,q_i';L,L_1}}{\dd L} = \Gamma \int\!\dd q_i\,
                            r\of{q_s,q_i,L} \E^{-\I \Delta k \left(L-2L_1\right)} \left[\tilde{\eta}^{\left(2\right)}\of{q_i,q_i';L,L_1}\right]^{*} \\
                        \displaystyle \frac{\dd\! \left[\tilde{\eta}^{\left(2\right)}\of{q_i,q_i';L,L_1}\right]^{*}}{\dd L} = 
                            \Gamma \int\!\dd q_s\,r^{*}\of{q_s,q_i,L} \E^{\I \Delta k \left(L-2L_1\right)}\beta^{\left(2\right)}\of{q_s,q_i';L,L_1} \end{cases}\\
                   \begin{aligned}
                        \left[\tilde{\eta}^{\left(2\right)}\of{q_i,q_i';L_1,L_1}\right]^{*} &= \delta\of{q_i-q_i'} \\
                        \beta^{\left(2\right)}\of{q_s,q_i';L_1,L_1} &= 0.
                   \end{aligned}
                \end{gathered}
            \end{equation}
            Assuming that 
            \begin{align}\label{eq:symm_prop_r_comp}
                r\of{q_s,q_i,2L_1-\El} &= r\of{q_s,q_i,\El},
            \end{align}
            which means that $r$ is mirror-symmetric around~$L_1$,
            it is possible to make the substitution ${\El=2L_1-L}$
            so that after redefining
            \begin{subequations}
                \begin{align}
                    \tilde{\eta}^{\left(2\right)}\of{q_i,q_i';2L_1-\El,L_1} &\eqdef \tilde{\Eta}^{\left(2\right)}\of{q_i,q_i';\El,L_1}, \\
                    \beta^{\left(2\right)}\of{q_s,q_i';2L_1-\El,L_1} &\eqdef \Beta^{\left(2\right)}\of{q_s,q_i';\El,L_1},
                \end{align}
            \end{subequations}
            the system above reads
            \begin{equation}\label{eq:integ_diffeq_sys_HB_equiv}
                \begin{gathered}
                    \int_{L_1}^{0}\!\dd \El\begin{cases}
                        \displaystyle \frac{\dd \Beta^{\left(2\right)}\of{q_s,q_i';\El,L_1}}{\dd \El} = -\Gamma \int\!\dd q_i\,
                            r\of{q_s,q_i,\El} \E^{\I \Delta k \El} \left[\tilde{\Eta}^{\left(2\right)}\of{q_i,q_i';\El,L_1}\right]^{*} \\
                        \displaystyle \frac{\dd\!\left[\tilde{\Eta}^{\left(2\right)}\of{q_i,q_i';\El,L_1}\right]^{*}}{\dd \El} = 
                            -\Gamma \int\!\dd q_s\,
                        r^{*}\of{q_s,q_i,\El} \E^{-\I \Delta k \El}\Beta^{\left(2\right)}\of{q_s,q_i';\El,L_1} \end{cases}\\
                   \begin{aligned}
                        \left[\Eta^{\left(2\right)}\of{q_i,q_i';L_1,L_1}\right]^{*} &= \delta\of{q_i-q_i'} \\
                        \Beta^{\left(2\right)}\of{q_s,q_i';L_1,L_1} &= 0.
                   \end{aligned}
                \end{gathered}
            \end{equation}
        \end{widetext}
        Obviously, the systems of integro-differential equations
        in~\eqref{eq:integ_diffeq_sys_HB_equiv}
        and~\eqref{eq:system_single_crystal} are equivalent.
        Generally, this would not have been the
        case if we had assumed that the systems are driven
        by different functions~$r^{\left(1\right)}$
        and~$r^{\left(2\right)}$.
        \par From that, we can identify the following properties relating the transfer
        functions of the first and the second crystal:
        \begin{subequations}
            \begin{align}
                \left[\tilde{\eta}^{\left(2\right)}\of{q,q';2L_1,L_1}\right]^{*} &= \tilde{\eta}^{\left(1\right)}\of{q',q;L_1,0},
                \label{eq:etabeta12_relation_L_nophase_eta}\\
                \beta^{\left(2\right)}\of{q,q';2L_1,L_1} &= \beta^{\left(1\right)}\of{q',q;L_1,0}.
                \label{eq:etabeta12_relation_L_nophase_beta}
            \end{align}
        \end{subequations}
        Note that these solutions still do not include
        the interferometer phase. When including it,
        the second relation instead reads
        \begin{align}\label{eq:beta_relation_L_phase}
            \beta^{\left(2\right)}\of{q,q';2L_1,L_1} &=
                \E^{\I\phi}\beta^{\left(1\right)}\of{q',q;L_1,0}.
        \end{align}
        Alternatively, to describe an \suoo interferometer,
        \Cref{eq:composite_tf_eta_phase,eq:composite_tf_beta_phase} can be used
        together with
        \Cref{eq:etabeta12_relation_L_nophase_eta,eq:etabeta12_relation_L_nophase_beta}.
        \par In this work, we have assumed a perfect compensation of the
        quadratic phase of the light, and therefore, we have only considered an
        interferometer phase which has no dependence on
        the transverse signal and idler wave vectors~$q_s$
        and~$q_i$. Generally however, when the compensation is not
        performed, that is, when the phase is $q$-dependent, 
        \Cref{eq:etabeta12_relation_L_nophase_eta,eq:etabeta12_relation_L_nophase_beta,eq:beta_relation_L_phase} do not hold. Nevertheless,
        in systems with a complicated $q$-dependent phase, as 
        long as it is possible to perform the compensation using a set of 
        lenses, mirrors or spatial light modulators (SLMs), the 
        necessary symmetry is induced and, therefore,
        \Cref{eq:etabeta12_relation_L_nophase_eta,eq:etabeta12_relation_L_nophase_beta,eq:beta_relation_L_phase} hold. In that case however,
        proper attention must be given to the
        losses introduced by these additional optical
        components since they reduce the phase
        sensitivity~\cite{Manceau2017}.

    \section{Covariances for a plane-wave and finite-width pump}
        \subsection{Plane-wave pump}\label{sec:cov_plots_appendix_pw}
\begin{figure*}[ptbh]%
                \subfloat{\label{fig:pw_var_ncom_0}%
                    \includegraphics[width=0.46\linewidth]
                        {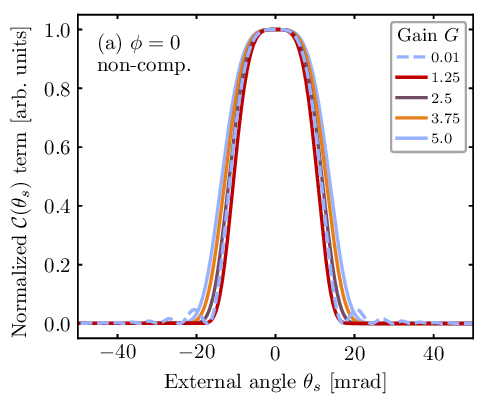}}%
                \subfloat{\label{fig:pw_var_ncom_05}%
                    \includegraphics[width=0.46\linewidth]
                        {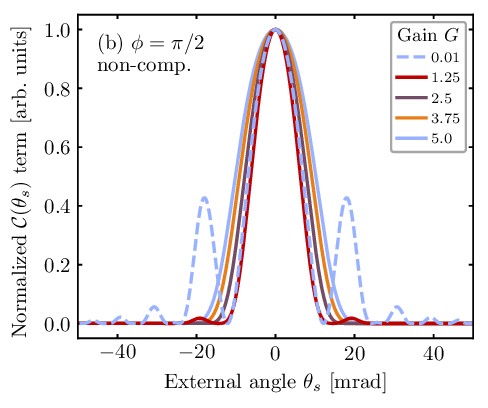}}\\
                \subfloat{\label{fig:pw_var_ncom_1}%
                    \includegraphics[width=0.46\linewidth]
                        {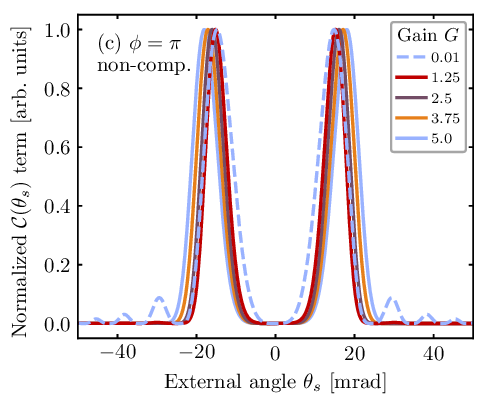}}%
                \subfloat{\label{fig:pw_var_comp_0}%
                    \includegraphics[width=0.46\linewidth]
                        {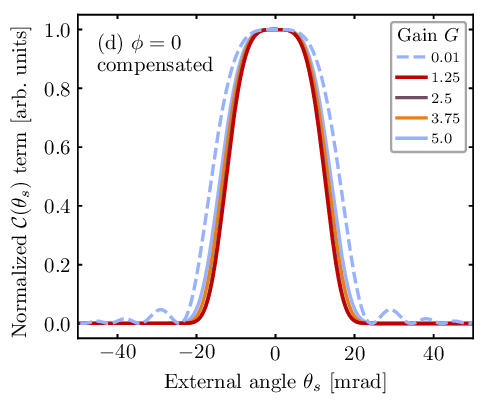}}\\%
                \subfloat{\label{fig:pw_var_comp_05}%
                    \includegraphics[width=0.46\linewidth]
                        {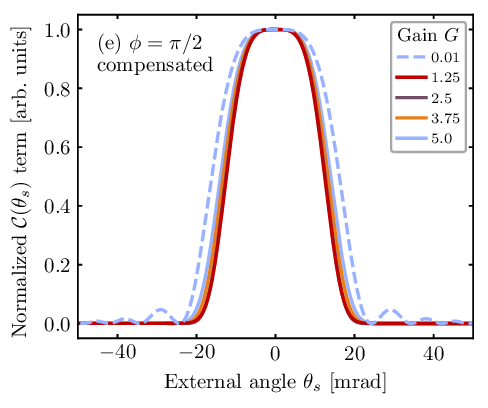}}%
                \subfloat{\label{fig:pw_var_comp_0995}%
                    \includegraphics[width=0.46\linewidth]
                        {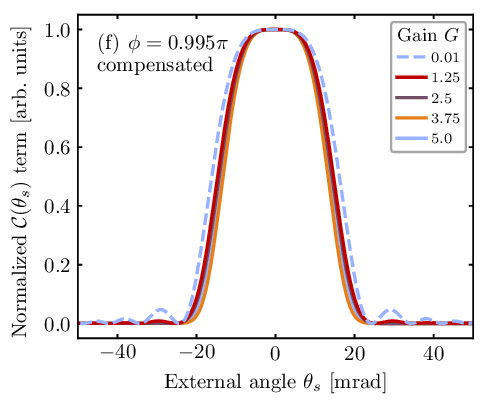}}%
                \caption{The finite term $\CC\of{\theta_s}$ of the
                    plane-wave covariance given by
                    \Cref{eq:pw_cov_general_finite_spectral} for 
                    the (a)--(c)~non-compensated interferometer and
                    for the (d)--(e)~compensated interferometer for different
                    phases. Note that for the compensated setup, the third phase we show
                    is not ${\phi=\pi}$ since then the covariance would be identically zero,
                    see \Cref{eq:intens_pw_su11_comp_N,eq:intens_pw_su11_comp_xi,eq:cov_pw_cc,eq:pw_cov_general_finite_spectral}.
                    \label{fig:pw_var}}%
            \end{figure*}

           \Cref{fig:pw_var_ncom_0,fig:pw_var_ncom_05,fig:pw_var_ncom_1,%
           fig:pw_var_comp_0,fig:pw_var_comp_05,fig:pw_var_comp_0995} 
           present the covariance profiles for different gains in the 
           plane-wave pump case. The profiles are described by the
           function~$\CC\of{q_s}$ defined in
           \Cref{eq:pw_cov_general_finite_spectral}.
           As one can see, the plane-wave covariance
           $\cov_{\mathrm{pw}}\of{q_s,q_s'}$
           [\Cref{eq:cov_pw_cc}]
           vanishes unless ${q_s=q_s'}$,
           therefore, the function $\CC$ contains all the information about
           the covariance. This is a direct consequence of the plane-wave
           pumping which results in perfect correlations between
           the plane-wave modes. As was already mentioned in the
           main text (\Cref{sec:pw_intro}), the covariance profiles
           calculated for the non-compensated setup
           [\Cref{fig:pw_var_ncom_0,fig:pw_var_ncom_05,fig:pw_var_ncom_1}], and
           for the compensated setup
           [\Cref{fig:pw_var_comp_0,fig:pw_var_comp_05,fig:pw_var_comp_0995}], 
           have shapes similar to the intensity profiles and broaden
           as the gain increases. Note that the figures are plotted in 
           the external angles.
        
        \subsection{Finite-width pump}\label{sec:cov_plots_appendix_fw}
\begin{figure*}[ptbh]%
                \subfloat[]{\label{fig:fw_cov_ncom}%
                    \includegraphics[width=0.925\linewidth]
                        {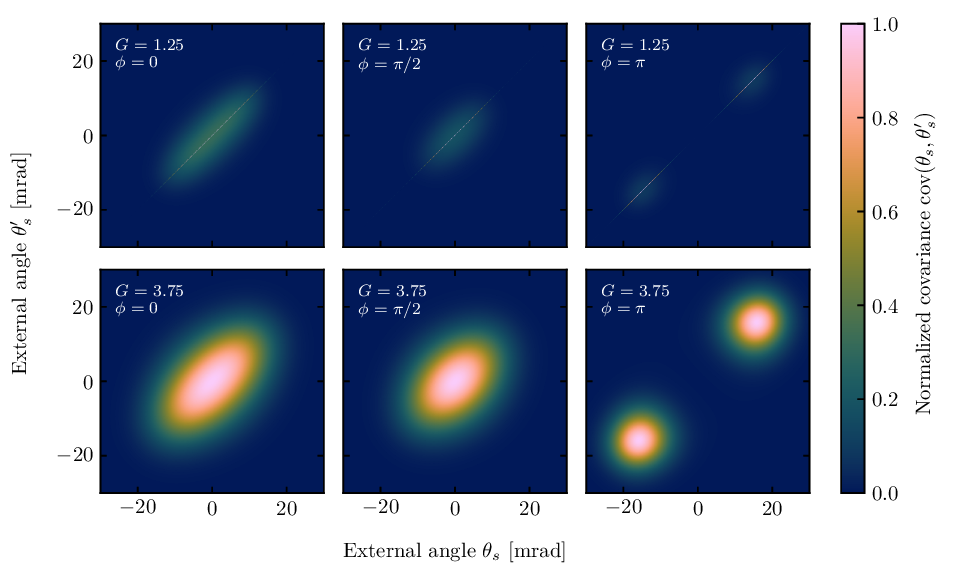}%
                }\\%
                \subfloat[]{\label{fig:fw_cov_comp}%
                    \includegraphics[width=0.925\linewidth]
                        {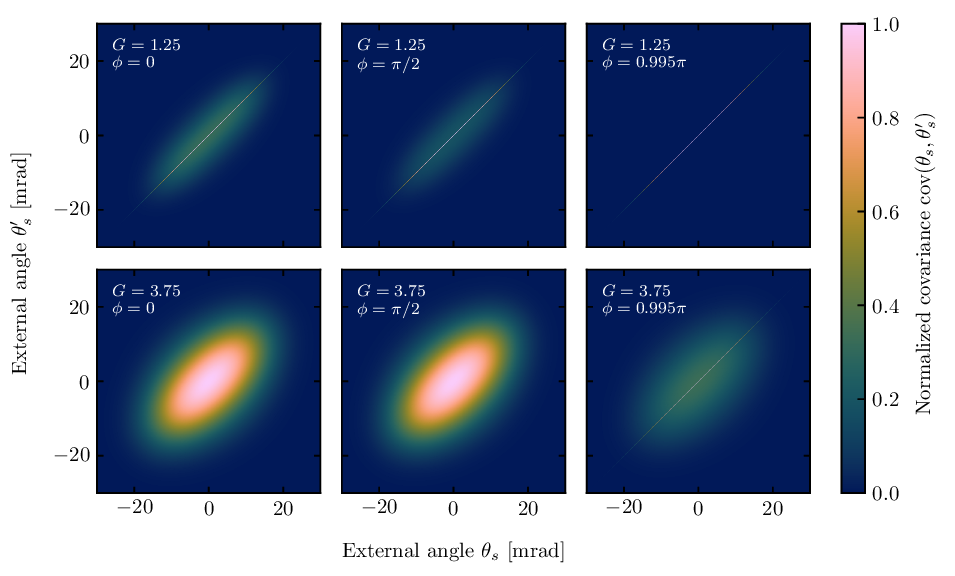}%
                }%
                \caption{Plots of the covariances $\cov\of{\theta_s,\theta_s'}$ for the
                    finite-width pump with an
                    intensity distribution FWHM of \SI{50}{\micro\meter}, 
                    (a)~for the non-compensated setup and (b)~for the
                    compensated setup, for several gains and phases. Note that each subplot
                    is normalized independently, so that
                    ${\max_{\theta_s,\theta_s'}\cov\of{\theta_s,\theta_s'}=1}$.
                    Furthermore, for the compensated setup, the third phase we show
                    is not ${\phi=\pi}$ since then the covariance would include only the shot noise term,
                    see \Cref{eq:cov_fw_phase_separated}.
                    \linebreak[3] This figure uses the ``Batlow'' color map~[F.~Crameri,
                    \href{https://doi.org/10.5281/zenodo.5501399}{%
                    \textit{Scientific colour maps (7.0.1)}} (Zenodo, 2021)].}%
            \end{figure*}

            Contrary to the plane-wave pump case,
            the covariance profiles for a finite-width pump are in general
            only fully described by considering the entire range for both
            arguments of the covariance function $\cov\of{q_s,q_s'}$.
            The plots for the non-compensated setup are shown in \Cref{fig:fw_cov_ncom}
            and for the compensated setup in \Cref{fig:fw_cov_comp}.
            Since we have to show these plots as colormesh-plots, we restrict ourselves to
            only showing two values for the parametric gain for each setup. Nevertheless,
            important features of the covariance functions become apparent.
            \par First, it should be noted that for both setups and for the
            small gain ${G=1.25}$,
            a sharp line can be observed along the diagonal. This feature
            is the shot-noise term 
            ${\delta\of{q_s-q_s'}\langle\hat{N}_s\of{q_s}\rangle}$ appearing 
            in \Cref{eq:cov_finwidth}. Numerically, the Dirac-delta is expressed
            via $\delta\of{x-y}\equiv\delta_{i_x,i_y}/\Delta x$, where~$i_x$ and~$i_y$
            are the indices on the numerical grid corresponding to the points~$x$ and~$y$,
            respectively, and~$\Delta x$ is the step size of the grid. Therefore,
            in the discretized case, we observe a line of single-pixel width
            and finite value along the diagonal. As the parametric gain increases,
            this term becomes
            negligible~\cite{PhysRevA.102.053725}, which results
            in the ellipse-shaped covariance for the larger gain value
            ${G=3.75}$. However, due to the perfect destructive
            interference in the compensated setup, the shot noise term becomes visible
            again as ${\phi\to\pi}$. This is due to the fact that the
            term resulting in the ellipse-shaped contribution scales
            as~${\sim\cos^4\of{\phi/2}}$ with the
            interferometer phase~$\phi$, while the shot noise
            term scales as~${\sim\cos^2\of{\phi/2}}$,
            see \Cref{eq:cov_fw_phase_separated} and \Cref{eq:intens_su11_fw_N,eq:xi_def_beta_eta},
            respectively.
            \par For the non-compensated setup, the covariance-ellipse narrows
            along the diagonal as the phase increases. Then, similar to
            the intensity profiles shown in \Cref{fig:fw_intens}, 
            a splitting of the covariance
            is observed as ${\phi\to\pi}$ which is caused by the
            imperfect destructive interference.
            For the compensated setup this is not the case. Instead, the widths
            of the ellipse along the diagonal and anti-diagonal stay unchanged.
            However, as mentioned above, the shot noise term becomes visible
            for phases close to $\pi$. Clearly, the ellipse shape is determined
            by the integral term in \Cref{eq:cov_fw_phase_separated} and
            is therefore independent of~$\phi$.
    
    \FloatBarrier
    \bibliography{bibliography}
    
\end{document}